\documentclass{IEEE-Con-Sys-mag}

\jvol{XX}
\jnum{XX}
\paper{8}
\jmonth{June}
\jname{IEEE CONTROL SYSTEMS}
\pubyear{2020}

\usepackage{amsmath,amssymb,amsfonts}
\usepackage{algorithmic}
\usepackage{graphicx}
\usepackage{textcomp}
\usepackage[font={small}]{caption}
\usepackage[compress]{cite}
\usepackage{dsfont}
\usepackage{epsfig}
\usepackage{enumerate}
\usepackage{graphics}
\usepackage{subcaption}
\usepackage{caption}
\usepackage{multirow}
\usepackage{framed}
\usepackage{hyperref}

\def\bs{\boldsymbol}
\def\mcal{\mathcal}
\def\mbb{\mathbb}

\def\mcx{\mathcal{X}}
\def\mcy{\mathcal{Y}}
\def\mca{\mathcal{A}}

\def\bfs{\mathcal{C}} % battlefield set

\def \be {\begin{equation}}
\def \ee {\end{equation}}
\def \ba {\begin{aligned}}
\def \ea {\end{aligned}}
\mathchardef\mhyphen="2D

\def\WL{\text{WL}}
\def\BS{\text{BS}}
\def\GL{\text{GL}}
\def\yes{{\color{green}\checkmark}}
\def\no{{\color{red}$\times$}}

\begin{document}

\title{Move Over, Prisoner's Dilemma\stitle{Colonel Blotto has arrived}}

\author{{K}eith Paarporn and Jason R. Marden}
\affil{}

\maketitle

\dois{}{}

\chapterinitial{A}n industrial control system operator monitoring a distributed manufacturing network faces a design challenge that transcends simple optimization: twenty-five process control loops, fifteen intrusion detection sensors, and potential vulnerabilities across SCADA communication channels. Which control loops merit redundant monitoring? How much sensing capacity suffices at each subsystem? This resource allocation problem becomes truly complex when the system's security depends not just on the administrator's choices, but on an adversary simultaneously solving the inverse problem, identifying which vulnerabilities to exploit given their own resource constraints. Similar strategic interdependence appears when the Coast Guard positions vessels to interdict maritime trafficking, when grid operators allocate reserves to maintain stability against potential disruptions, or when wildlife agencies deploy rangers against adaptive poaching networks \cite{pita2008deployed,tambe2011security,yang2014adaptive,an2012protect}.

\begin{figure*}[!ht]
    \centering
    \includegraphics[width=0.95\linewidth]{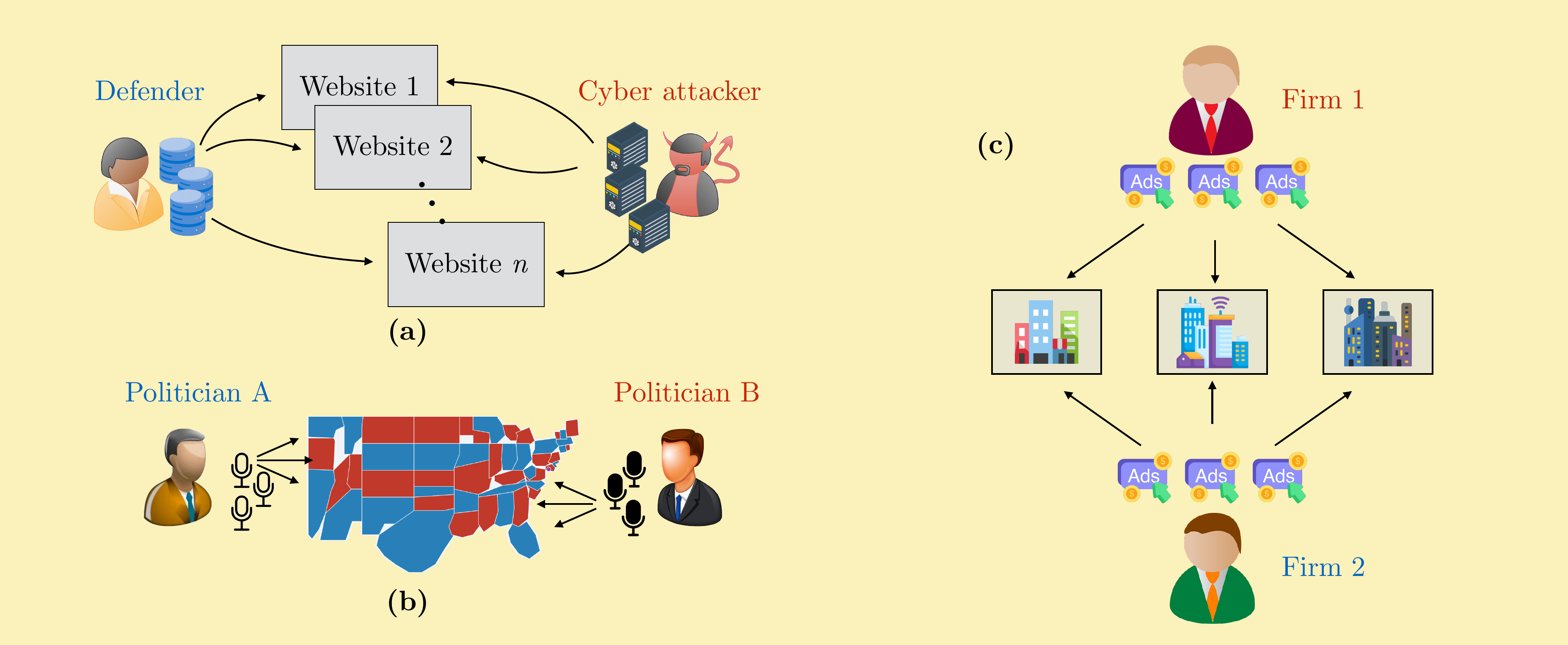}
    \caption{
    % Three examples of competitive resource allocation captured by the Colonel Blotto framework: adversarial interaction, strategic decision-making, and limited resources distributed across a collection of contests. (a) A network defender allocates security resources across websites against a cyber attacker. (b) Two rival politicians allocate campaign resources across districts. (c) Two competing firms bid fixed advertising budgets across multiple simultaneous online ad auctions.
    Three examples of competitive resource allocation for which the Colonel Blotto game captures the essential common elements -- adversarial interaction, strategic decision-making, and limited resources to utilize across a collection of contests. (a) A defender of a cyber system is responsible for preventing cyber attacks on a collection of websites \cite{chia2011colonel,Ferdowsi_2017,iliaev2023tullock,atkinson2023resource}. The defender allocates technological resources such as computing power, security software, and infrastructure to protect the system. The cyber attacker allocates computing resources in an effort to compromise as many websites as possible. (b) Two rival politicians compete in an election \cite{merolla2004lotto,kovenock2008electoral,Behnezhad_2018}. Each must decide how to allocate campaigning resources such as awareness advertisements, speaking engagements, and rallies. (c) Two competing firms place advertisements across multiple geographic regions in order to maximize market share.}
    \label{fig:examples}
\end{figure*}

These scenarios exemplify a class of control problems where the value of an allocation strategy cannot be evaluated in isolation. The effectiveness of deploying state estimation resources to one subsystem over another depends fundamentally on where an attacker concentrates their disruption efforts, which in turn depends on anticipated monitoring strategies. This interdependence, where optimal control actions must account for an intelligent, optimizing adversary, necessitates game-theoretic analysis rather than classical optimization alone.

Historically, each application domain developed its own models and solution techniques. Cybersecurity researchers constructed defender-attacker frameworks for network security \cite{do2017game,cardenas2008secure,cardenas2009challenges,cavusoglu2008security,pinar2010optimization}. Economists built market competition models for advertising allocation \cite{zia2019search,li2025online}. Political scientists devised campaign resource models \cite{lake1979new,gurian1986resource}. While each community made progress within its domain, this fragmentation carried costs.  Researchers repeatedly derived similar structural insights without recognizing underlying commonalities. Practitioners faced problems that didn't quite match their domain's existing models had nowhere to turn. Most critically, the absence of a unifying framework meant that fundamental insights about competitive resource allocation, insights that transcend specific applications, remained scattered across disconnected literatures.

The need for a more unified approach becomes clear when we recognize that these diverse problems share essential structure: multiple simultaneous contests, limited resources, strategic interdependence, and the requirement to balance competing objectives. What varies between domains are surface details, terminology, context, and specific constraints, not the underlying strategic considerations. This recognition motivates the search for a general framework that captures common elements while remaining flexible enough to accommodate domain-specific features.

\begin{pullquote}
	The Prisoner's Dilemma, zero-sum games, LQR team problems, and differential games have shaped game theory in controls for decades, but the field's most pressing adversarial challenges demand a richer framework, and its name is Colonel Blotto.
\end{pullquote}

\begin{summary}
\summaryinitial{T}he Prisoner's Dilemma, zero-sum games, LQR team problems, and differential games have shaped game theory in controls for decades, but the field's most pressing adversarial challenges demand a richer framework, and its name is Colonel Blotto.
Strategic adversarial constraints represent a fundamental consideration in control systems, from cybersecurity defense to infrastructure protection.
Colonel Blotto games, despite their direct relevance to such applications, remain underutilized in the controls community relative to other game-theoretic approaches. This article aims to close that gap for the controls community.  Indeed, theoretical advances within the last two decades have spurred a resurgence of interest and enabled their applications across several domains.

In this article, we introduce the Colonel Blotto framework, survey 
key analytical and computational results, and demonstrate 
how problems spanning cybersecurity, network defense, and 
multi-agent systems fit naturally within this structure. 
Three research directions are examined in depth: interdependent 
contest objectives that capture networked vulnerabilities, 
alternate winning rules that model partial rewards and 
structural asymmetries, and multi-agent competitive 
environments involving coalition formation and strategic 
concessions. Taken together, these directions reveal a 
framework that is both practically deployable and rich enough 
to capture the strategic complexity inherent in adversarial 
resource allocation.
\end{summary}

\begin{pullquote}
	The goal of this article is to showcase the value of the Colonel Blotto framework to the controls community.
\end{pullquote}

% \begin{pullquote}
% 	The very complexity that has historically resisted clean solutions ``makes Blotto all the more compelling in its interpretations".
% \end{pullquote}

Colonel Blotto games provide precisely this framework. 
Indeed, Figure~\ref{fig:examples} illustrates three seemingly disparate real-world examples that share a common underlying structure that the Colonel Blotto framework is able to accommodate.
These models, dating back to 1921 \cite{Borel}, abstract competitive resource allocation to its essential elements: players with limited resources, allocations across multiple contests, and payoffs determined by strategic decision-making. 
Despite their direct relevance to strategic resource allocation, Blotto games remain underutilized in the controls community relative to other game-theoretic models such as the Prisoner's Dilemma.
The disparity stems largely from analytical complexity. In a Prisoner's Dilemma game, it suffices to model just two possible strategies, cooperate or defect. In the Colonel Blotto game, there is a continuum of allocation strategies. This, coupled with the competitive nature of the interaction (which usually entails randomized strategies), often has made the analysis of Blotto games inaccessible.
Nevertheless, the very complexity that has historically resisted clean solutions ``makes Blotto all the more compelling in its interpretations'' \cite{Golman_2009}.

\begin{figure*}[!ht]
    \centering
    \includegraphics[scale=0.25]{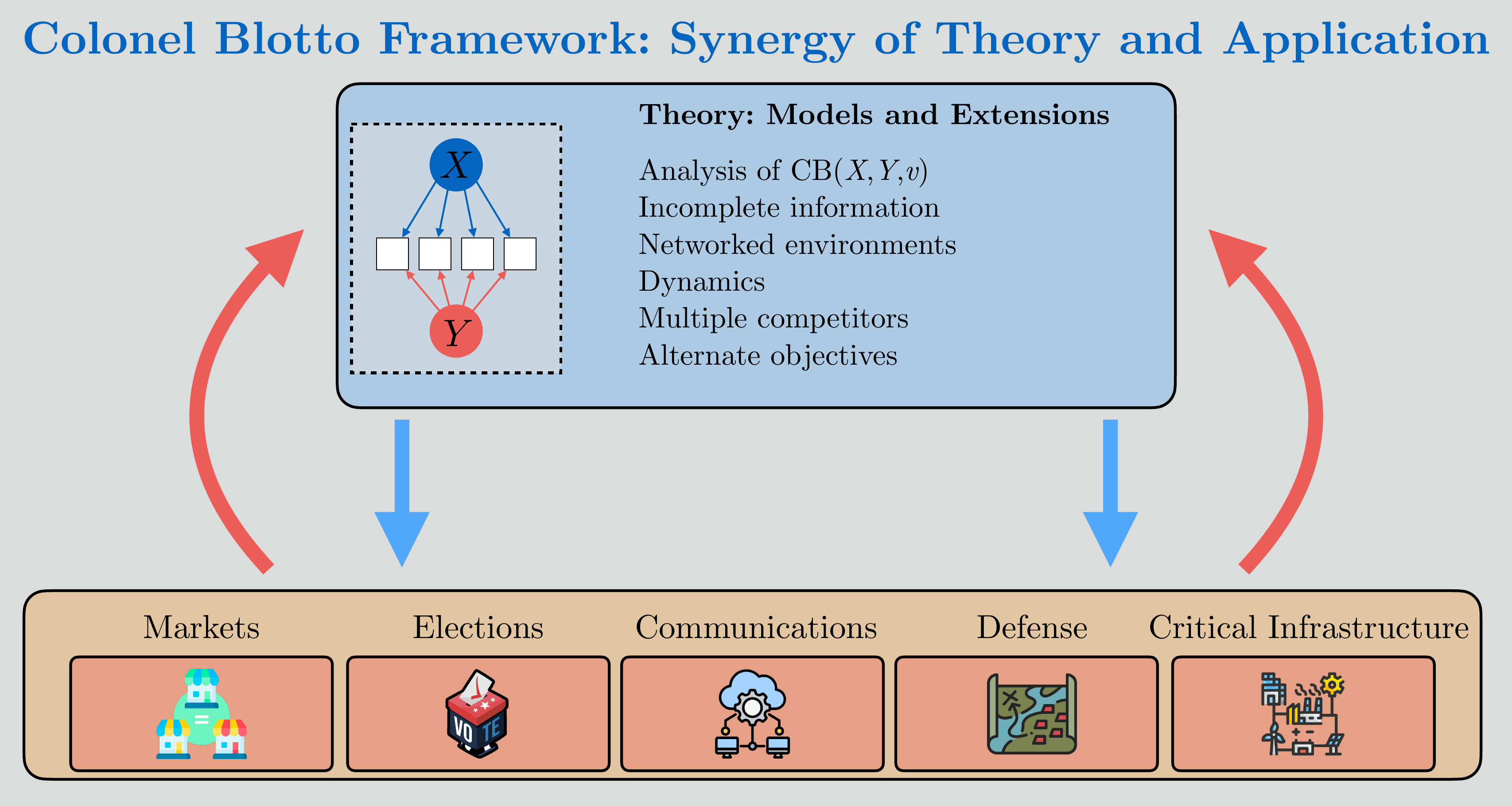}
    \caption{This figure highlights the synergy between applications and the emerging Colonel Blotto landscape.  The analysis of the classic Blotto games over several decades of research has given rise to numerous interesting research directions (\cite{Borel,Gross_1950,Roberson_2006,Hart_2008,Golman_2009,Schwartz_2014,Thomas_2018,kovenock2021generalizations,perchet2022algorithmic}). A non-exhaustive list of these directions include: 
    incomplete information (\cite{Adamo_2009,Fuchs_2012,Gupta_2014b,duffy2015stochastic,Kim_2017,Kovenock_2011,Paarporn_2019,Ewerhart_2021,Paarporn_2019,Paarporn_2022_LCSS,paarporn2024incomplete,brethouwer2024general,diaz2025value}), network effects (\cite{Shubik_1981,Shahrivar_2014,Kovenock_2018,cortes2018generalising,Guan_2019,Shishika_2021,aghajan2023extension,diaz2023beyond,aghajan2026defense}), dynamic decision-making (\cite{arad2012multi,vu2019combinatorial,Aidt_2019,leon2021bandit,ferguson2022ensuring,Shishika_2021}), and multi-player competitions (\cite{Adamo_2009,Boix_2020,diaz2023beyond,Kovenock_2012,paarporn2021division}).
    These fundamental extensions of the Blotto game highlight the utility of the theoretical framework to practitioners in numerous application domains (blue arrows), who have access to proven solutions with performance guarantees when real-world problems can be mapped into the Blotto framework. 
    These application domains include: market competitions and elections (\cite{lake1979new,Myerson_1993,kovenock2008electoral,kohli2012colonel,rogers2013colonel,washburn2013or,masucci2014strategic,Behnezhad_2018,Kovenock_2012}),  communication networks (\cite{abdelraheem2017cooperative,charatsaris2022competitive,halabi2023stealthy,Shahrivar_2014,wang2020privacy,wu2009optimal}), defensive operations (\cite{Shubik_1981,roberson2005colonel,powers2009colonel,collins2012colonel,baudains2016colonel,lamb2022benefits,grimsman2025colonel}), the security of critical infrastructure (\cite{Fuchs_2012,Ferdowsi_2017,guan2018colonel,Ferdowsi_2020,Shahrivar_2014}),  and cybersecurity (\cite{chia2011colonel,Gupta_2014b,min2017defense,Gupta_2014a,atkinson2023resource,halabi2023stealthy,iliaev2023tullock,paarporn2024preventive}).
    In turn, the need for principled decision-making in various applications spurs the continued theoretical development of the Colonel Blotto framework (red arrows). This feedback relationship between theory and applications has created an ongoing and thriving research agenda centered around Colonel Blotto games. Attribution: graphic icons from \texttt{flaticon.com}.
    }
    \label{fig:framework}
\end{figure*}

Recent theoretical developments over the past two decades have inspired renewed interest in Blotto games across several disparate disciplines, including systems and control. This resurgence has given rise to what we refer to as the \emph{Colonel Blotto framework}. The framework's impact manifests along two dimensions. First, it establishes a common analytical foundation for researchers. Rather than developing bespoke models for each new scenario, investigators can build upon established equilibrium characterizations, solution algorithms, and structural results. 
This shared language has enabled fundamental contributions spanning incomplete information, network effects, dynamic decision-making, and multi-player competitions. Second, practitioners gain access to proven solutions with performance guarantees. When a real-world allocation problem can be mapped into the Colonel Blotto framework, whether in infrastructure defense, communication security, or economic competition, decision-makers can immediately leverage decades of theoretical results. An adversarial resource allocation problem transforms from an open question into an instance of a well-characterized class, with accompanying solution methods and optimality guarantees.  This synergy is illustrated in Figure~\ref{fig:framework}.

\subsection{Article objectives and organization}

The goal of this article is to showcase the value of the Colonel Blotto framework to the controls community. We begin by introducing the classic formulation, its foundational equilibrium properties, and a historical account of the field's development from its origins in 1921 through Roberson's 2006 breakthrough to the General Lotto relaxation that has enabled much of the modern analysis. The subsequent sections examine three directions that have proven especially fruitful for security and resource allocation in engineered systems: interdependent contest objectives that capture networked vulnerabilities, alternate winning rules that model partial rewards and structural asymmetries, and multi-agent competitive environments involving coalition formation and strategic concessions. Throughout, we highlight how each direction translates into actionable insights for resource-constrained operators in adversarial settings.

\begin{pullquote}
	Recent theoretical developments over the past two decades have inspired renewed interest in Blotto games across several disparate disciplines, including systems and control.
\end{pullquote}

\section{The Colonel Blotto Game}\label{sec:cb}

A Colonel Blotto game models competitive resource allocation in its 
most essential form: two players with limited budgets simultaneously 
distribute resources across multiple contested objectives, with payoffs 
determined by relative allocations at each contest. Despite this 
apparent simplicity, the framework captures the strategic 
interdependence that characterizes a wide range of adversarial 
allocation problems relevant to control systems. 
%Figure~\ref{fig:examples} illustrates three seemingly disparate real-world examples that share a common underlying structure that the Colonel Blotto framework accommodates.

\subsection{The Model}

\begin{figure}[!ht]
    \centering
    \includegraphics[width=0.8\linewidth]{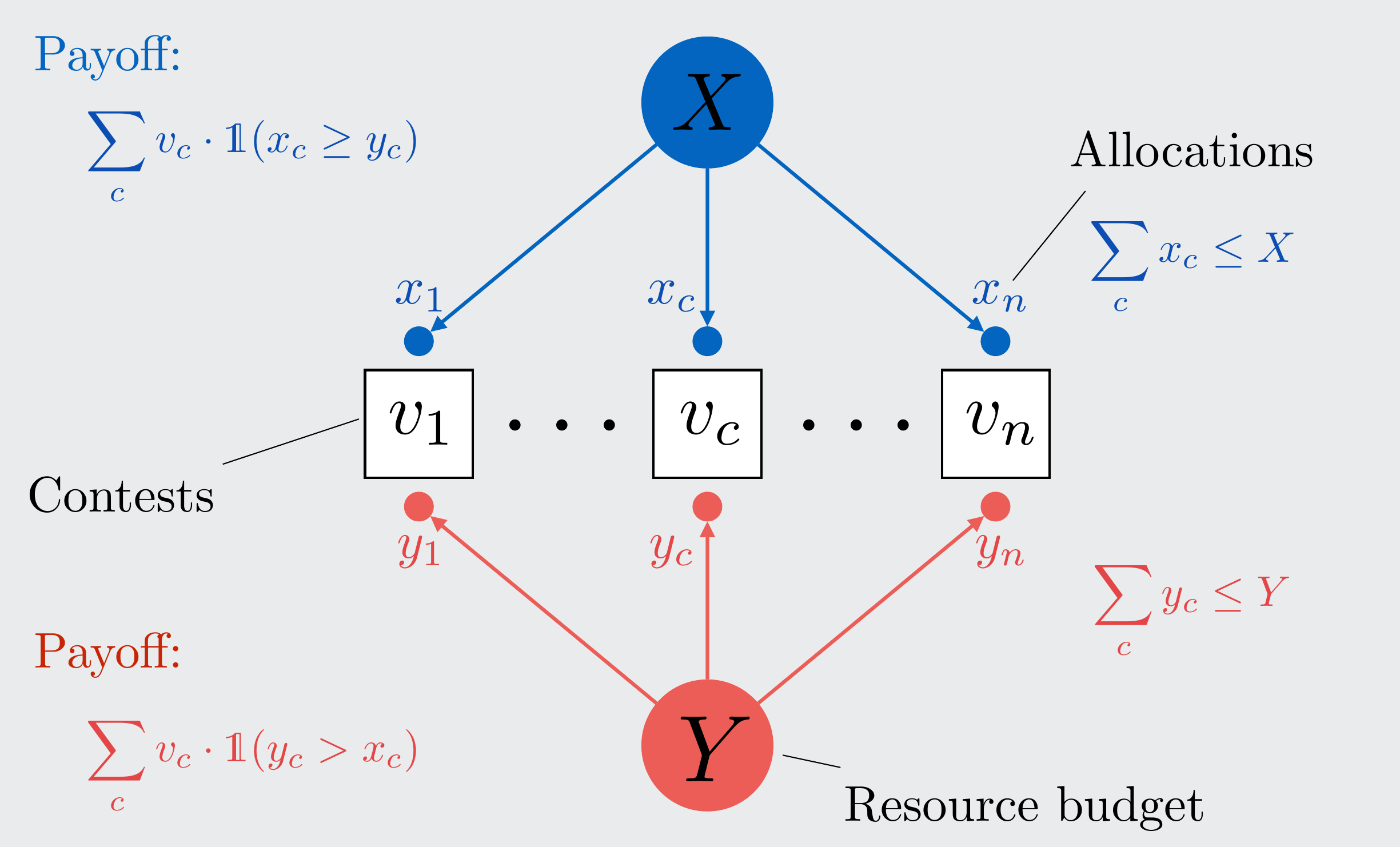}
    \caption{The basic components of a Colonel Blotto game. Two players, 
$\mcx$ and $\mcy$, compete over $n$ simultaneous contests with limited 
resource budgets $X, Y > 0$. A player secures a contest by 
outspending its opponent, and each player's overall payoff is the 
cumulative value of secured contests.}
    \label{fig:cb-game}
\end{figure}

Each of the scenarios above shares the same essential structure: two 
budget-constrained players simultaneously allocating resources across 
multiple contests, with outcomes at each contest determined by their 
relative investments. Formalizing this structure, a \emph{Colonel 
Blotto game} consists of two players $\mcx$ and $\mcy$ with fixed 
resource budgets $X > 0$ and $Y > 0$, competing over a set of $n$ 
contests labeled $\bfs := \{1,\ldots,n\}$. Each contest $c \in \bfs$ 
has an associated valuation $v_c > 0$ common to both players, 
collected in the vector $\bs{v} := (v_1,v_2,\ldots,v_n)$. A feasible 
allocation for player $\mcx$ is any vector $\bs{x} = (x_1,\ldots,x_n)$ 
belonging to the set
\be
    \mca(X) := \left\{ \bs{x} \in \mbb{R}^n : \sum_{c\in\bfs} x_c 
    \leq X, \ x_c \geq 0 \ \forall c\in\bfs \right\},
\ee
and similarly $\bs{y} \in \mca(Y)$ for player $\mcy$. A player secures a contest by allocating strictly more resources to it than its opponent, capturing the contest's full valuation. For simplicity, ties are awarded to player $\mcx$, a convention that can be relaxed without significant changes to the fundamental results \cite{kovenock2021generalizations}. Given an action profile $(\bs{x},\bs{y})$, the payoff to player $\mcx$ is
\be\label{eq:blotto_payoff}
    % u_\mcx(\bs{x},\bs{y}) := \sum_{c\in\bfs} v_c \cdot 
    % \left(\mathds{1}(x_c > y_c) + \frac{1}{2}\mathds{1}(x_c = y_c) 
    % \right),
    u_\mcx(\bs{x},\bs{y}) := \sum_{c\in\bfs} v_c \cdot 
    \mathds{1}(x_c \geq y_c),
\ee
where $\mathds{1}(\cdot)$ is the indicator function that takes value 1 if the condition is true, and 0 otherwise. The payoff to player $\mcy$ is determined as $u_\mcy(\bs{x},\bs{y}) = 
\phi - u_\mcx(\bs{x},\bs{y})$, where $\phi := \sum_{c\in\bfs} v_c$ is the aggregate value of contests. Because the payoffs are constant-sum ($u_\mcx(\bs{x},\bs{y}) + u_\mcy(\bs{x},\bs{y})  = \phi$), we refer only to $u_\mcx$ going forward, and denote this game as $\text{CB}(X,Y,\bs{v})$. While this ``winner-take-all" rule is the canonical choice for individual contest outcomes, richer models of competitive interaction are possible and are examined in this article.  A diagram of a Colonel Blotto game is provided in Figure~\ref{fig:cb-game}.

\begin{figure*}[!ht]
    \centering
    \includegraphics[scale=0.22]{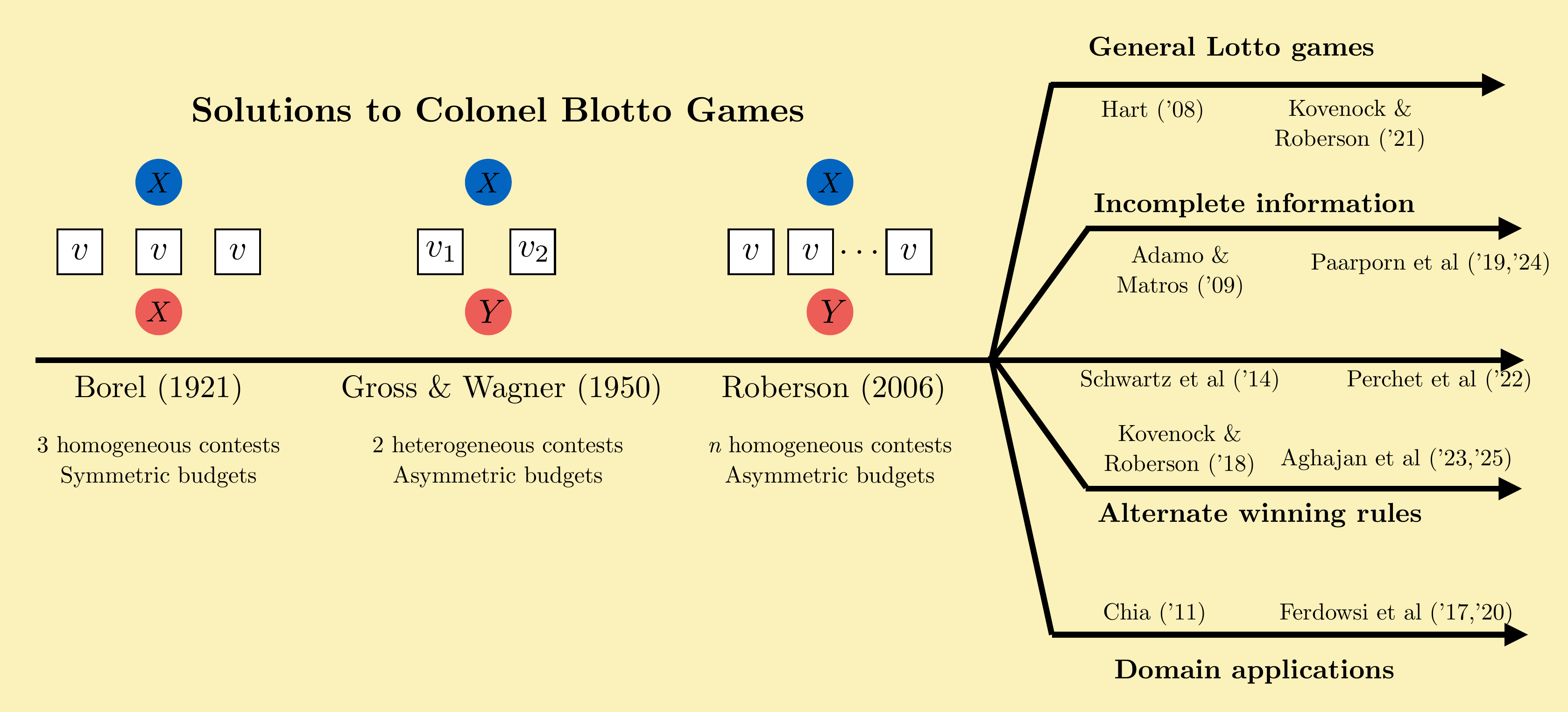}
    \caption{This figure depicts a non-exhaustive timeline regarding research progress made in the analysis of Blotto games and its variants over the last century. It is interesting to note that Gross and Wagner's solution \cite{Gross_1950} to all two-battlefield Blotto games remained the state-of-the-art for the entire latter half of the 20th century. Roberson's work in 2006 \cite{Roberson_2006}, which developed methods to analyze Blotto games with $n$ homogeneous contests, was a significant advance. It also marked the beginning of intensified research activity in this area, sprouting several parallel threads of study. For example, researchers often seek to apply these novel tools to generate domain-specific insights. Another thread considers other variants such as the General Lotto game, and researchers continue to further extend the generality of the results. This flurry of still ongoing activity is depicted in Figure~\ref{fig:framework}.}
    \label{fig:timeline}
\end{figure*}

\subsection{Equilibrium Analysis and Performance Guarantees}

Since Colonel Blotto games exhibit constant-sum structure, optimal 
behavior generally requires \emph{mixed strategies} -- probability 
distributions over pure allocations rather than deterministic 
commitments. Randomization is not merely a mathematical convenience; 
it reflects the practical reality that a deterministic allocation 
strategy is exploitable by any adversary who can anticipate it. We 
denote a mixed strategy for player $\mcx$ by $s_\mcx \in 
\Delta(\mca(X))$, representing a randomization over the feasible 
allocation set $\mca(X)$, and define the expected payoff as
\begin{eqnarray}\label{eq:EUi}
    U_\mcx(s_\mcx,s_\mcy) := \mbb{E}_{\substack{\bs{x} \sim s_\mcx 
    \\ \bs{y} \sim s_\mcy}} \left[ u_\mcx(\bs{x},\bs{y}) \right]. 
\end{eqnarray}
An \emph{equilibrium} of the Colonel Blotto game is a strategy profile 
$(s_\mcx^*,s_\mcy^*)$ from which neither player benefits by unilateral 
deviation, satisfying
\begin{equation}
    U_\mcx(s_\mcx,s_\mcy^*) \leq U_\mcx(s_\mcx^*,s_\mcy^*) \leq 
    U_\mcx(s_\mcx^*,s_\mcy),
\end{equation}
for all $s_\mcx \in \Delta(\mca(X))$ and $s_\mcy \in \Delta(\mca(Y))$. 
The left inequality states that $s_\mcy^*$ limits $\mcx$'s payoff to 
at most $U_\mcx(s_\mcx^*,s_\mcy^*)$ regardless of $\mcx$'s strategy, 
while the right inequality states that $s_\mcx^*$ guarantees $\mcx$ 
at least this same payoff regardless of $\mcy$'s strategy.

Von Neumann's minimax theorem \cite{v1928theorie} establishes that in two-player 
constant-sum games, the equilibrium payoff $V := 
U_\mcx(s_\mcx^*,s_\mcy^*)$ is unique across all equilibria, and may 
be computed via either formulation:
\begin{equation}\label{eq:maxmin_minmax}
    V = \max_{s_\mcx} \left\{ \min_{s_\mcy} U_\mcx(s_\mcx,s_\mcy) 
    \right\} = \min_{s_\mcy}\left\{ \max_{s_\mcx} 
    U_\mcx(s_\mcx,s_\mcy) \right\}.
\end{equation}
This duality has a direct parallel in control theory: $V$ plays the role of a worst-case performance bound, analogous to the $\mathcal{H}_\infty$ norm, and any equilibrium strategy functions as a \emph{security strategy} that guarantees this bound against all possible adversarial responses. Just as $\mathcal{H}_\infty$ synthesis seeks a controller that bounds the worst-case disturbance-to-output gain, equilibrium play in a Colonel Blotto game seeks an allocation that bounds the worst-case competitive loss, regardless of how the opponent allocates its resources.

The analysis of Colonel Blotto games has a long and rich history of incremental progress. 
See ``\nameref{sidebar-history}'' for more details on the landmark results over the last 100 years.
A visual timeline of this historical progress, extending into present-day research directions, is illustrated in Figure~\ref{fig:timeline}. In addition, Table \ref{table:blotto_papers} identifies numerous research works in recent years that have contributed fundamental extensions to the Colonel Blotto game.

\begin{sidebar}{A Century of Progress: From Borel to Roberson and Beyond}
\section[A Century of Progress: From Borel to Roberson and Beyond]{}\label{sidebar-history}

\setcounter{sequation}{0}
\renewcommand{\thesequation}{S\arabic{sequation}}
\setcounter{stable}{0}
\renewcommand{\thestable}{S\arabic{stable}}
\setcounter{sfigure}{0}
\renewcommand{\thesfigure}{S\arabic{sfigure}}

\sdbarinitial{T}he Colonel Blotto game has a storied history spanning more 
than a century, yet its most significant analytical advances 
have come only recently. Figure~\ref{fig:timeline} provides 
a graphical account of this progression. The first 
mathematical formulation appeared in \'{E}mile Borel's 1921 
work \cite{Borel_SA}, predating Von Neumann's minimax 
theorem by several years. Borel derived equilibrium strategies 
for two players with symmetric budgets ($X = Y$) competing 
over three equally valued battlefields, establishing that 
even in the simplest symmetric cases, optimal play requires 
sophisticated randomization over feasible allocations.

%\cite{v1928theorie}

The next significant advance came nearly three decades later. 
In a 1950 research memorandum to the US Air Force, Gross and 
Wagner characterized equilibrium strategies for asymmetric 
players ($X \neq Y$) over two battlefields of arbitrary value 
\cite{Gross_SA}. Their work introduced the now-standard 
``Colonel Blotto" terminology and attracted sustained interest 
among military strategists throughout the 1950s and 1960s 
\cite{blackett1954some,blackett1958pure,bellman1969colonel,
Shubik_1981}. Despite this early attention, research activity 
remained sparse for decades and Gross and Wagner's 
two-battlefield solution would stand as the state of the art 
for over fifty years.

\subsection{Roberson's breakthrough}

The field was transformed in 2006, when economist Brian 
Roberson developed methods to characterize equilibrium 
strategies for asymmetric players competing over three or 
more homogeneous battlefields \cite{Roberson_SA}. The 
primary technical challenge was finding equilibrium mixed 
strategies with support over the $n$-dimensional simplex, a 
problem that becomes substantially harder beyond two 
battlefields as the marginal distributions across contests 
can no longer be determined independently. Roberson's 
breakthrough triggered a wave of applied research, as 
practitioners across cybersecurity, infrastructure 
protection, and economic competition recognized that 
multi-contest resource allocation lies at the heart of their 
problems \cite{Gupta_2014a,ferdowsi2017colonel,Ferdowsi_2020}. 
Simultaneously, the constraints of these applications 
accelerated continued theoretical development. 
Figure~\ref{fig:framework} illustrates this productive 
feedback between theory and application.

A central development in this expanding framework has been 
the \emph{General Lotto game} 
\cite{Hart_SA,kovenock_2021_SA}, which relaxes the 
budget constraint to hold only in expectation rather than 
with probability one. This 
seemingly modest modification dramatically simplifies 
equilibrium characterization while preserving essential 
strategic features and admitting closed-form solutions for 
arbitrary parameter configurations, a feat that remains 
elusive for classical Colonel Blotto games except in special 
cases. The General Lotto formulation has become a preferred 
foundation for analyzing increasingly complex scenarios~\cite{Myerson_1993}, 
several of which are examined in the sections that follow.

\subsection{The expanding research landscape}

The research activity that followed Roberson's 2006 results catalyzed a rapid expansion 
of the Colonel Blotto framework across multiple dimensions. 
Table~\ref{table:blotto_papers} provides a systematic 
overview of this literature, organizing recent contributions 
along several key axes that characterize model complexity 
and applicability. Research has progressed along multiple 
frontiers: relaxing the two-player assumption to accommodate 
coalitional dynamics and multi-agent competitions; 
introducing incomplete or asymmetric information structures; 
incorporating network effects and interdependent objectives; 
addressing discrete resource constraints relevant to 
real-world implementations; and developing sequential and 
dynamic formulations that capture temporal decision-making. 
The literature spans multiple disciplines: economics, 
control theory, computer science, and operations research, 
each bringing distinct methodological approaches and 
application perspectives. The subsequent sections examine 
three of these directions in depth, selected for their 
particular relevance to security and resource allocation 
problems in engineered systems.

\end{sidebar}

\begin{sidebar}{Analytical Foundations: The General Lotto Game}
\section[Analytical Foundations: The General Lotto Game]{}\label{sidebar-lotto}

\setcounter{sequation}{0}
\renewcommand{\thesequation}{S\arabic{sequation}}
\setcounter{stable}{0}
\renewcommand{\thestable}{S\arabic{stable}}
\setcounter{sfigure}{0}
\renewcommand{\thesfigure}{S\arabic{sfigure}}

The \emph{General Lotto game} is defined 
identically to the Colonel Blotto game  $\text{CB}(X,Y,\bs{v})$ with one exception: 
the budget constraint need only hold in expectation rather 
than with probability one. Specifically, admissible strategies 
$s_\mcx$ now belong to the set $\mcal{L}(X)$, defined as
\begin{equation}\label{eq:lotto_constraint}
    \mcal{L}(X) := \left\{ s_\mcx \in \Delta(\mbb{R}^n_+) : 
    \mbb{E}_{\bs{x} \sim s_\mcx}\left[\sum_{c\in\bfs} x_c
    \right] \leq X \right\}.
\end{equation}
Under this relaxation, player $\mcx$ may randomize over 
allocations that individually exceed the budget 
($\sum_{c\in\bfs} x_c > X$), as long as the expected 
expenditure remains within budget. Similar relaxations hold for player $\mcy$. 
Figure~\ref{sfig-lotto} illustrates the geometric distinction: Colonel Blotto 
strategies must have support confined to the simplex 
$\mca(X)$, whereas General Lotto strategies may place 
probability mass beyond this boundary while satisfying the 
expectation constraint. This modification allows one to decouple the 
marginal distributions across contests, dramatically 
simplifying equilibrium analysis while preserving 
fundamental strategic trade-offs.

\sdbarfig{\includegraphics[width=18.0pc]{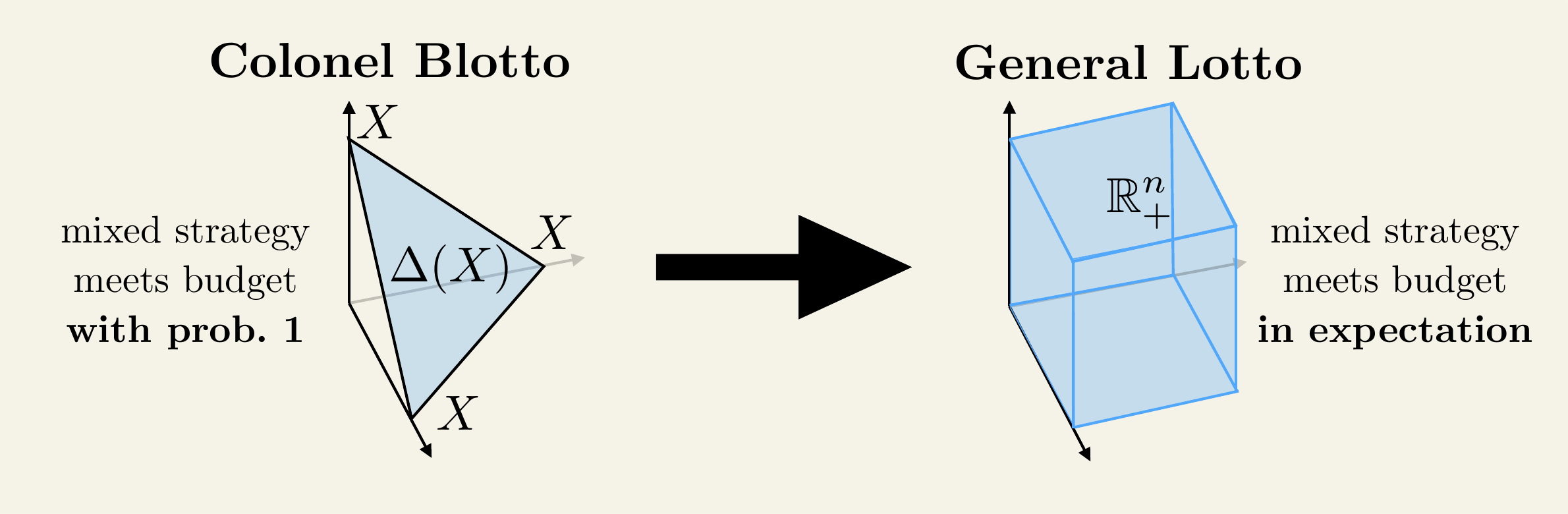}}{In Colonel Blotto games, the support of a mixed strategy is confined to the simplex (scaled with respect to the player's budget). In General Lotto games, the admissible strategies are relaxed such that the budget need only be met in expectation.\label{sfig-lotto}}

%\begin{figure}[!ht]
%    \centering
%    \includegraphics[scale=0.2]{figs/lotto.pdf}
%    \caption{The admissible space of mixed strategies for 
%    Colonel Blotto games and General Lotto games. In Colonel 
%    Blotto games, the support of a mixed strategy is confined 
%    to the simplex (scaled with respect to the player's budget). In 
%    General Lotto games, the admissible strategies are relaxed 
%    such that the budget need only be met in expectation.}
%    \label{fig:lotto_strategies}
%\end{figure}

The key benefit of this relaxation is analytical tractability: 
unlike $\text{CB}(X,Y,\bs{v})$, the General Lotto game admits 
complete equilibrium characterizations for arbitrary parameter 
configurations. The following result, established across 
multiple works \cite{Hart_SA_2,kovenock_2021_SA}, provides closed-form 
expressions.

\begin{theorem}\label{thm:GL}
   Consider an instance of the General Lotto game. Then in any equilibrium, the payoff to player $\mcx$ is
    \begin{equation}
        \phi \cdot L(X,Y),
    \end{equation}
    where
    \begin{equation}\label{eq:lotto_equil}
        L(X,Y) :=
        \begin{cases}
            \frac{X}{2Y}, &\text{if } X < Y \\[2mm]
            1-\frac{Y}{2X}, &\text{if } X \geq Y 
        \end{cases},
    \end{equation}
    and $\phi := \sum_{c\in\bfs} v_c$. The payoff to player 
    $\mcy$ is $\phi(1 - L(X,Y))$.\end{theorem}

The clean characterization of this result explains the General 
Lotto game's widespread adoption as the analytical foundation 
for numerous extensions in the Blotto framework.
As the Nobel Award-winning economist Roger Myerson stated in 1993, ``the advantage of [this] simplified formulation is that it will enable us to go beyond this `Colonel Blotto' literature and get results about more complicated situations" \cite{myerson_SA}.
%the extensions examined in the sections that follow

Two features of this result are worth highlighting. First, 
equilibrium payoffs depend only on the budget ratio $X/Y$ 
and the aggregate contest values $\phi$. Neither the 
number of contests nor their individual valuations affects 
the outcome, which stands in stark contrast to Colonel Blotto 
games where heterogeneous valuations substantially complicate 
analysis.  Hence, we will express an instance of a General Lotto game as $\text{GL}(X,Y,\phi)$. Second, $L(X,Y)$ is monotone in the expected 
direction: payoffs increase linearly with one's own resources 
and decrease hyperbolically with the opponent's, reflecting 
the competitive nature of the interaction.

The relationship between General Lotto and Colonel Blotto 
payoffs is systematic and well-understood. In symmetric 
scenarios ($X = Y$), both formulations yield identical 
equilibrium payoffs of $\phi/2$. For asymmetric budgets, the 
resource-advantaged player receives a lower payoff in General 
Lotto than in Colonel Blotto, while the resource-disadvantaged 
player receives a higher payoff, as the relaxed constraint 
affords the weaker player additional strategic flexibility. 
Critically, as the number of battlefields grows large 
($n \to \infty$), the two formulations' payoffs converge 
\cite{kovenock_2021_SA}, and for moderately large 
battlefield sets (for example, $n \geq 10$) General Lotto provides 
an accurate approximation of classical Colonel Blotto 
equilibria. Figure~\ref{sfig-CBvsGL} illustrates this 
relationship.

\sdbarfig{\includegraphics[width=12.0pc]{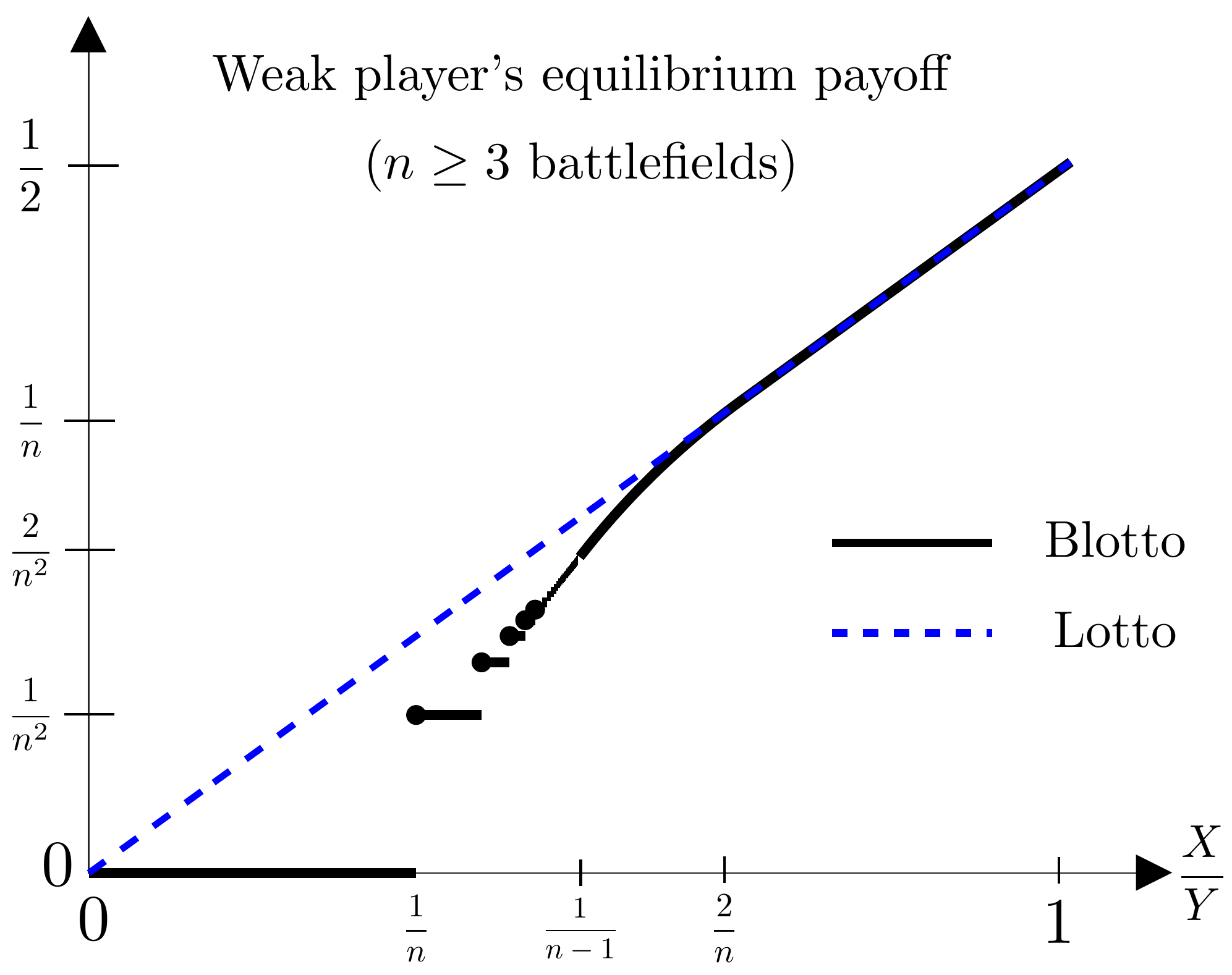}}{The solid black line indicates the equilibrium payoff to the weaker player $\mcx$ 
    ($X \leq Y$) in $\text{CB}(X,Y, (1/n)\mathds{1}_n)$ (solid black line), the Blotto game with $n$ equally-valued 
    contests at $1/n$. It is 
    piecewise constant for budget ratios $X/Y < \frac{1}{n-1}$ 
    with a countably infinite number of 
    discontinuities. We refer the interested reader to \cite{Roberson_2006} for the full details. The dashed blue line indicates the corresponding payoff in the General Lotto game, 
    $\text{GL}(X,Y,1)$.  The two lines coincide for $\frac{2}{n} 
    \leq X/Y \leq 1$. In the limit of large $n$, the Colonel 
    Blotto equilibrium payoff converges pointwise to that of the General Lotto game.\label{sfig-CBvsGL}}

%\begin{figure}[!ht]
%    \centering
%    \includegraphics[scale=0.2]{figs/CBvsGL.pdf}
%    \caption{The solid black line indicates the equilibrium payoff to the weaker player $\mcx$ 
%    ($X \leq Y$) in $\text{CB}(X,Y, (1/n)\mathds{1}_n)$ (solid black line), the Blotto game with $n$ equally-valued 
%    contests at $1/n$. It is 
%    piecewise constant for budget ratios $X/Y < \frac{1}{n-1}$ 
%    with a countably infinite number of 
%    discontinuities. We refer the interested reader to \cite{Roberson_2006} for the full details. The dashed blue line indicates the corresponding payoff in the General Lotto game, 
%    $\text{GL}(X,Y,1)$.  The two lines coincide for $\frac{2}{n} 
%    \leq X/Y \leq 1$. In the limit of large $n$, the Colonel 
%    Blotto equilibrium payoff converges pointwise to that of the General Lotto game.}
%    \label{fig:CBvsGL}
%\end{figure}

\end{sidebar}

\section{Highlighted Results}\label{sec:h-results}

The Colonel Blotto framework's value lies not only in the 
classic formulation $\text{CB}(X,Y,\bs{v})$, but in the 
rich collection of extensions it supports. 
We examine three directions that have proven especially 
fruitful for security and resource allocation in engineered 
systems. The first extends the classic linear-count objective 
to interdependent contest structures, capturing the cascading 
vulnerabilities and all-or-nothing outcomes prevalent in 
networked control problems. The second broadens the class of 
admissible winning rules through contest success functions, 
enabling richer models of partial rewards, stochastic 
outcomes, and structural asymmetries between competitors. 
The third moves beyond the two-player setting to examine 
multi-agent competitive environments, where coalition 
formation, strategic concessions, and informational 
mechanisms give rise to a qualitatively richer class of 
strategic considerations. 

Across all three directions, we 
emphasize how the resulting equilibrium characterizations 
translate into actionable insights for resource-constrained 
operators in adversarial settings.
Before proceeding, we first note that the foundations of much of this recent research is built upon the \emph{General Lotto game}, a popular variant of the original Blotto game. We direct the interested reader to ``\nameref{sidebar-lotto}'' for more details.

\begin{table*}[!ht]
    \begin{center}
        \begin{tabular}{|c|c|c|c|c|c|c|c|c|c|c|c|c|c|c|}
        \hline
        Paper & \multicolumn{3}{|c|}{\# Players} & \multicolumn{3}{|c|}{Resource type} & & \multicolumn{2}{|c|}{Information} & \multicolumn{3}{|c|}{\# Stages} & \multicolumn{2}{|c|}{Contest type}  \\    
        \hline
        & \rotatebox{90}{2-player} & \rotatebox{90}{3-player} & \rotatebox{90}{$n$-player} & \rotatebox{90}{Continuous} & \rotatebox{90}{Discrete} & \rotatebox{90}{Multi-type} & \rotatebox{90}{Symmetric} & \rotatebox{90}{Complete} & \rotatebox{90}{Incomplete} & \rotatebox{90}{One-shot} & \rotatebox{90}{2 or 3-stage} & \rotatebox{90}{$N$-stage} & \rotatebox{90}{Linear-count} & \rotatebox{90}{Interdependent}   \\
        \hline
        Classic Setting
        \multirow{2}{*}{} &  &  &  &  &  &  &  &  &  & &  &  &  &  \\
        \cite{Gross_1950,Roberson_2006,Kvasov_2007,Schwartz_2014,Thomas_2018,kovenock2021generalizations,perchet2022algorithmic} & \yes & \no & \no & \yes & \no & \no & \yes & \yes & \no & \yes & \no & \no & \yes & \no  \\
        \hline
        Incomplete information
        \multirow{3}{*}{} &  &  &  &  &  &  &  &  &  & &  &  &  &  \\
        \cite{Kovenock_2011,Fuchs_2012,Kim_2017,Paarporn_2019,Paarporn_2022_LCSS,paarporn2024incomplete,diaz2025value} & \yes & \no & \no & \yes & \no & \no & \yes & \no & \yes & \yes & \no & \no & \yes & \no  \\
        \cite{Adamo_2009,Kim_2017,Ewerhart_2021} & \no & \no & \yes & \yes & \no & \no & \yes & \no & \yes & \yes & \no & \no & \yes & \no  \\
        \hline  
        Interdependent \& alternate winning rules 
        \multirow{2}{*}{} &  &  &  &  &  &  &  &  &  & &  &  &  &  \\
        \cite{Shubik_1981,Golman_2009,Kovenock_2018,chandan2022strategic,aghajan2023extension,aghajan2023equilibrium} & \yes & \no & \no & \yes & \no & \no & \yes & \yes & \no & \yes & \no & \no & \no & \yes  \\
        \hline
        Multi-agent conflicts
        \multirow{2}{*}{} &  &  &  &  &  &  &  &  &  & &  &  &  &  \\
        \cite{Kovenock_2012,Gupta_2014a,Gupta_2014b,heyman2018colonel,Chandan_2020,paarporn2024strategically,diaz2023beyond} & \no & \yes & \yes & \yes & \no & \no & \no & \yes & \no & \no & \yes & \no & \yes & \no \\
        \hline
        Network defense 
        \multirow{2}{*}{} &  &  &  &  &  &  &  &  &  & &  &  &  &  \\
        \cite{Shahrivar_2014,Kovenock_2018,Guan_2019,Shishika_2021,aghajan2023extension,aghajanNetworkedCDC2022} & \yes & \no & \no & \yes & \no & \no & \yes & \yes & \no & \yes & \no & \no & \no & \yes  \\
        \hline
        Multiple resource types 
        \multirow{2}{*}{} &  &  &  &  &  &  &  &  &  & &  &  &  &  \\
        \cite{Vu_EC2021,chandan2022strategic,aghajan2023multiresource} & \yes & \no & \no & \yes & \no & \yes & \yes & \yes & \no & \yes & \yes & \no & \yes & \no  \\
        \hline
        Integer allocations 
        \multirow{2}{*}{} &  &  &  &  &  &  &  &  &  & &  &  &  &  \\
        \cite{Behnezhad_2017,Behnezhad_2018,Ahmadinejad_2019,dehghani2021computational} & \yes & \no & \no & \no & \yes & \no & \yes & \yes & \no & \yes & \no & \no & \yes & \no  \\
        \hline
        Sequential moves
        \multirow{2}{*}{} &  &  &  &  &  &  &  &  &  & &  &  &  &  \\
        \cite{Aidt_2019,Shishika_2021,chen2023path} & \yes & \no & \no & \no & \yes & \no & \yes & \no & \yes & \no & \no & \yes & \yes & \no  \\
        \hline
        Reinforcement learning 
        \multirow{2}{*}{} &  &  &  &  &  &  &  &  &  & &  &  &  &  \\
        \cite{vu2019combinatorial,leon2021bandit,beaglehole2023sampling,valles2024fast} & \yes & \no & \no & \no & \yes & \no & \yes & \no & \yes & \no & \no & \yes & \yes & \no  \\
        \hline
        \end{tabular}
    \end{center}
    \caption{This table compiles a large yet non-exhaustive literature on Blotto games and its variants that followed Roberson's 2006 paper. The papers are categorized into several major themes that mark fundamental and significant extensions to the original, classic Colonel Blotto game $\text{CB}(X,Y,\bs{v})$. Each of these themes can be mapped into one or more of listed directions from the discussion of ``Underlying assumptions and their limitations". The columns provide indicate distinct strategic features that can be modeled with Blotto games. For example, the rows ``Interdependent winning rules" and ``Network defense" reflect scenarios where player payoff functions cannot be described by the linear-count objective \eqref{eq:blotto_payoff}. The contributions to this literature come from multiple disciplines, including control systems and engineering, economics, optimization, and computer science fields. }
    \label{table:blotto_papers}
\end{table*}

\section*{\textbf{Direction 1: From Independent to 
Interdependent Contests}}

Classical formulations of Colonel Blotto and General Lotto games assume players 
seek to maximize cumulative value across independent contests \eqref{eq:blotto_payoff}.
This form of payoff is known as a \emph{linear-count objective} \cite{Kovenock_handbook_2012}. 
The allocation to and outcome of one contest exerts no influence on that of any other contest,
an assumption that proves inadequate for modeling the 
cascading effects and systemic vulnerabilities prevalent in 
networked control problems. An attacker breaching a computer 
network need only exploit a single vulnerability to compromise 
the entire system, while defenders must secure every attack 
surface to maintain integrity. Similarly, in power grid 
operations, the failure of critical transmission lines can 
trigger cascading failures regardless of robust defenses 
elsewhere in the network.

% \cite{pinar2010optimization}

Such scenarios motivate examination of interdependent 
objective structures that capture these all-or-nothing 
outcomes. We focus on two canonical forms representing 
opposite extremes: \emph{weakest-link} objectives, where 
system integrity depends on defending every attack surface, 
and \emph{best-shot} objectives, where exploiting any single 
vulnerability suffices for system breach. These formulations 
naturally model defender-attacker scenarios in cybersecurity, 
infrastructure protection, and defensive operations where 
asymmetric winning conditions fundamentally shape strategic 
behavior.

We model these scenarios using a General Lotto 
formulation, denoting the defender as $\mcx$ and the attacker 
as $\mcy$, each with fixed resource budgets $X > 0$ and 
$Y > 0$, competing over a set of $n$ vulnerabilities $\bfs$. 
The weakest-link objective for the defender is defined by
\be\label{eq:WL_objective}
    u_\WL(\bs{x},\bs{y}) := \mathds{1}\left\{ x_c \geq y_c 
    \ \text{for all } c\in\bfs \right\}
\ee
and the best-shot objective for the attacker is defined by
\be\label{eq:BS_objective}
    u_\BS(\bs{x},\bs{y}) := \mathds{1}\left\{ y_c > x_c 
    \ \text{for some } c\in\bfs \right\},
\ee
where $\bs{x},\bs{y}$ denote the players' resource 
allocations across vulnerabilities. These objectives 
establish a constant-sum structure: $u_\BS(\bs{x},\bs{y}) 
= 1 - u_\WL(\bs{x},\bs{y})$. Unlike the linear-count 
objective \eqref{eq:blotto_payoff}, individual 
vulnerabilities carry no inherent valuations; the system 
exists in one of two states, breached or secure, with the 
outcome determined by relative resource allocations across 
all vulnerabilities simultaneously. The structural asymmetry 
inherently favors the attacker: exploiting any single 
vulnerability suffices to compromise the system, while 
defense requires universal coverage. We denote this 
formulation as $\text{WL}(X,Y,\bfs)$, with feasible 
strategies $s_\mcx \in \mcal{L}(X)$ and $s_\mcy \in 
\mcal{L}(Y)$ as in the General Lotto game (see ``\nameref{sidebar-lotto}''). 
The expected 
payoffs $U_\WL(s_\mcx,s_\mcy)$ and $U_\BS(s_\mcx,s_\mcy)$ 
thus represent the probability of maintaining security and the 
probability of system breach, respectively.

\begin{figure}
    \centering
    \includegraphics[width=1.0\linewidth]{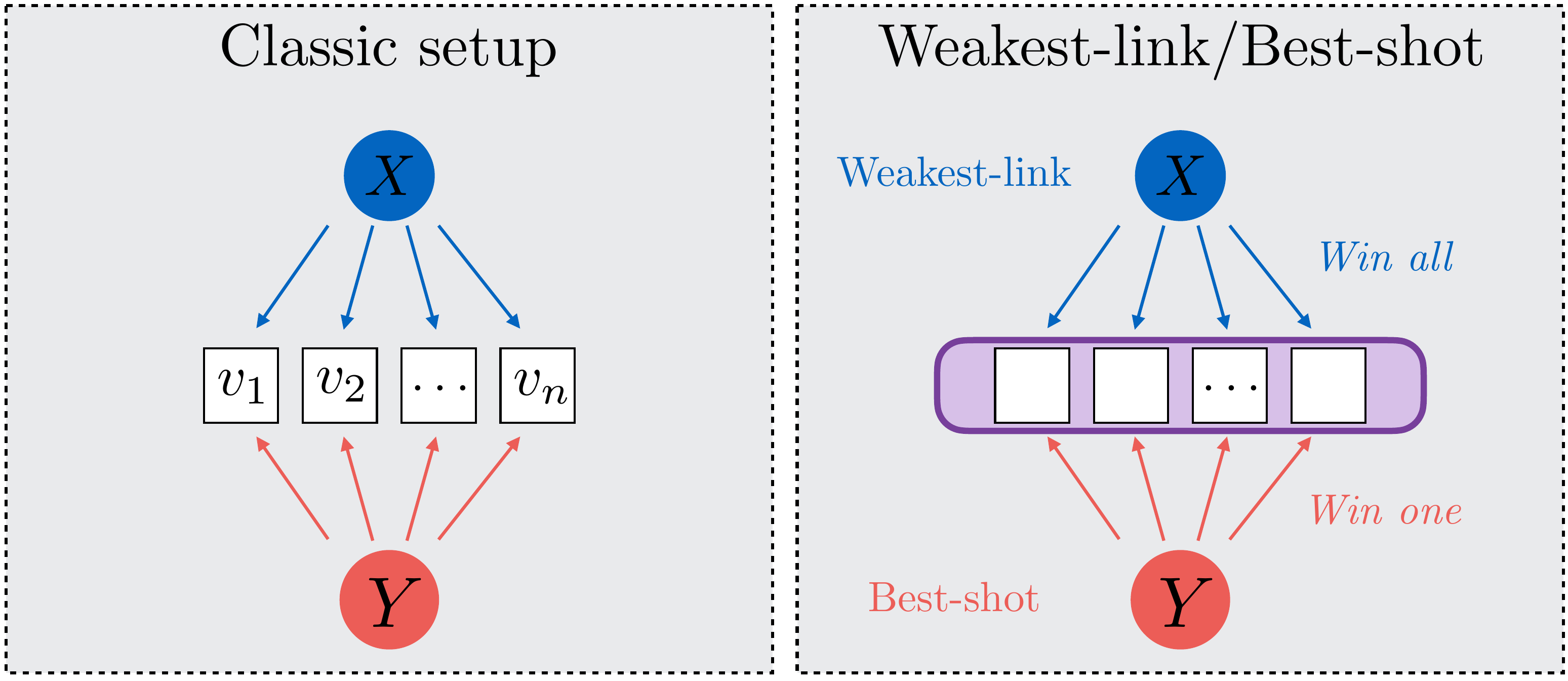}
    \caption{(Left) The classic setup of 
    $\text{CB}(X,Y,\bs{v})$ or $\text{GL}(X,Y,\bs{v})$, in 
    which both players seek to maximize cumulative value by 
    securing individual battlefields. (Right) The 
    weakest-link/best-shot formulation $\text{WL}(X,Y,\bfs)$. 
    The defender $\mcx$ must secure all $n$ vulnerabilities 
    to succeed; the attacker $\mcy$ needs only one. Individual 
    vulnerabilities carry no inherent valuations, as the 
    binary outcome depends on allocations across all 
    vulnerabilities simultaneously.}
    \label{fig:WLBS}
\end{figure}

The following theorem provides the equilibrium 
characterization for this defender-attacker scenario.

\begin{theorem}[Adapted from \cite{Kovenock_2018,aghajan2023extension}]
\label{thm:WLBS}
    Consider the weakest-link/best-shot formulation 
    $\text{WL}(X,Y,\bfs)$. In an equilibrium, the payoff to 
    the defender ($\mcx$) is given by
    \begin{equation}\label{eq:pi_WL}
        \pi_{\WL}(X,Y,\bfs) := 
        \begin{cases}
            \frac{X}{2|\bfs|Y}, &\text{if } X < |\bfs|\cdot Y 
            \\[2mm]
            1 - \frac{|\bfs| Y}{2X}, &\text{if } X \geq 
            |\bfs|\cdot Y
        \end{cases}
    \end{equation}
    and the payoff to the attacker ($\mcy$) is given by 
    $\pi_{\BS}(X,Y,\bfs) := 1 - \pi_{\WL}(X,Y,\bfs)$.
\end{theorem}

\begin{sidebar}{Beyond Weakest-Link: Objectives with Networked Interdependence}
\section[Beyond Weakest-Link: Objectives with Networked Interdependence]{}\label{sidebar-interdependent}

\setcounter{sequation}{0}
\renewcommand{\thesequation}{S\arabic{sequation}}
\setcounter{stable}{0}
\renewcommand{\thestable}{S\arabic{stable}}
\setcounter{sfigure}{2}
\renewcommand{\thesfigure}{S\arabic{sfigure}}

\sdbarinitial{T}he weakest-link and best-shot objectives, defined in~\eqref{eq:WL_objective} and~\eqref{eq:BS_objective}, reflect an extreme form of contest interdependence in which every contest is equally exposed, capturing the very essence of attacker-defender scenarios.  
Here, we present a variety of other formulations involving more localized interdependence structures, where rewards are based on securing particular subsets of the contests.
This flexibility demonstrates how the framework can be sharply tailored to domain-specific objectives.

\subsection{Network path defense}

In a network path defense problem, the defender must secure all nodes along at least one path from a source to destination node, forming a ``Best-shot of Weakest-link" objective. This objective can express, for example, the need to ensure a communication line from a sender to receiver, or to secure a path for transport in a logistics chain. On the other hand, the attacker seeks to block every available path by compromising at least one node on every path, forming a ``Weakest-link of Best-shot" objective. This setup is illustrated in Figure~\ref{sfig-pathdefense}.

These composite objectives are built directly from the modular building blocks of Theorem~\ref{thm:WLBS}, illustrating how more complex networked defense scenarios can be constructed and analyzed within the same framework. We refer the reader to the papers~\cite{Kovenock_2018_SA,aghajan_2024_SA}, which leverage these building blocks to synthesize solutions to a large family of attack-defense interactions.

\sdbarfig{\includegraphics[width=14.0pc]{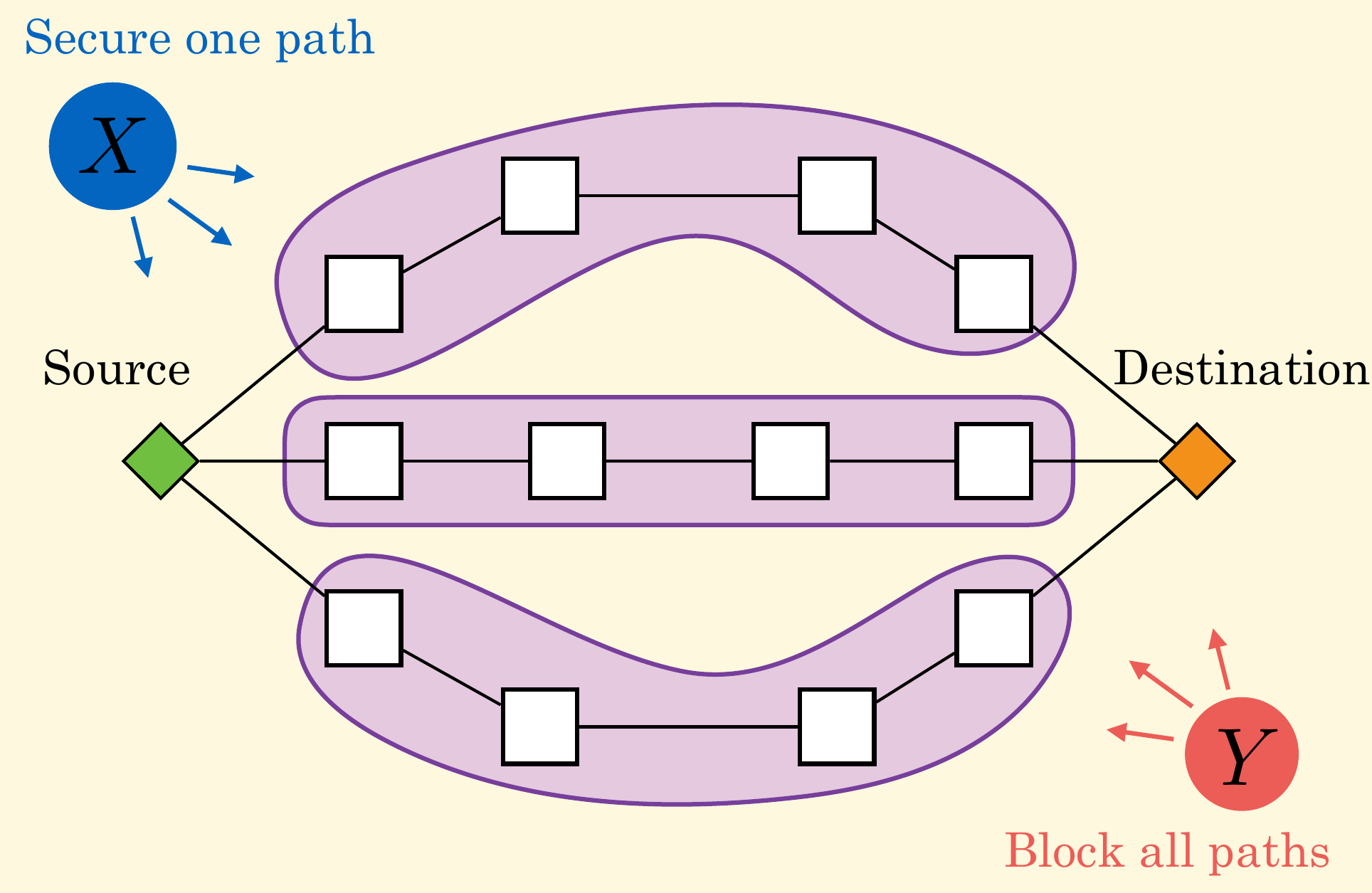}}{A network path defense problem. In this diagram, there are three possible paths from source to destination, and there are four nodes along each of the paths.\label{sfig-pathdefense}}

\subsection{The Majoritarian objective}

Another form of contest interdependence involves the requirement to secure any subset of contests that constitutes a majority in order to win the prize. The majoritarian objective is defined as
\begin{sequation}
    u_{maj}(\bs{x},\bs{y}) := \mathds{1}\left\{ x_c \geq y_c 
    \ \text{for a majority of } c\in\bfs \right\}.
\end{sequation}
This type of objective is naturally applicable to political elections. Indeed, many works have studied resource allocation strategies for certain variants of the majoritarian objective, specifically in the context of the US electoral college \cite{dehghani_2021_SA}.
These studies are primarily algorithmic in nature, owing to the fact that equilibrium strategies for majoritarian objectives have only been analytically characterized for three contests~\cite{szentes2003beyond,aghajan2023equilibrium}.

\subsection{Preserving network structures}

Several works examine networked objectives concerned 
with forming or preserving desired graph-theoretic properties.
For example,~\cite{shahrivar_2014_SA} formulates a network formation game in which players submit bids to purchase possible edges of a graph.
The resulting set of edges owned by a player determines its resulting payoff.
Other work extends the attacker-defender dichotomy by linking the defender's objective to the preservation of certain graph structures, such as the network's connectivity, the average path length, or the average node's degree \cite{guan_2019_SA}.
In a similar vein,~\cite{aghajan_2026_SA} focuses on deriving precise equilibrium strategies when the defender's objective is to preserve as many edges in a network as possible.

\subsection{Multi-resource extensions}

The Blotto framework primarily considers  allocation decisions in which the deployed resources represent only a single type of asset. For example, they may represent the allocation of only finances, only employees, or only troops. However, in many competitive interactions, multiple distinct types of resources are often deployed in efforts to gain the upper hand. This warrants multi-resource extensions where the winner of a contest is determined not only by the amount of resources allocated, but also by the composition of resource types that are allocated~\cite{aghajan_2023_SA,lamb_2022_SA}. Under this context, networked effects play a role. Some resources can be more effective in competing against other types, which inform allocation strategies that are aware to these networked relationships.

\end{sidebar}

Comparing \eqref{eq:pi_WL} with the General Lotto payoff 
\eqref{eq:lotto_equil} reveals a precise structural 
connection. The defender's equilibrium payoff in 
$\text{WL}(X,Y,\bfs)$ coincides with its payoff in the 
General Lotto game $\GL(X,|\bfs|\cdot Y, 1)$, that is, as 
if it faced an opponent with $|\bfs|$ times its original 
budget. The equilibrium strategies themselves differ 
markedly from those in the classic General Lotto game: the 
defender allocates the same randomly drawn amount to every 
vulnerability, consistent with its objective to achieve 
universal coverage, while the attacker concentrates all 
resources on a single randomly selected vulnerability, 
consistent with its objective to breach just one. For the interested reader, complete 
descriptions of these strategies appear in~\cite{Kovenock_2018,
aghajan2023extension}.

\begin{figure}
    \centering
    \includegraphics[width=0.6\linewidth]{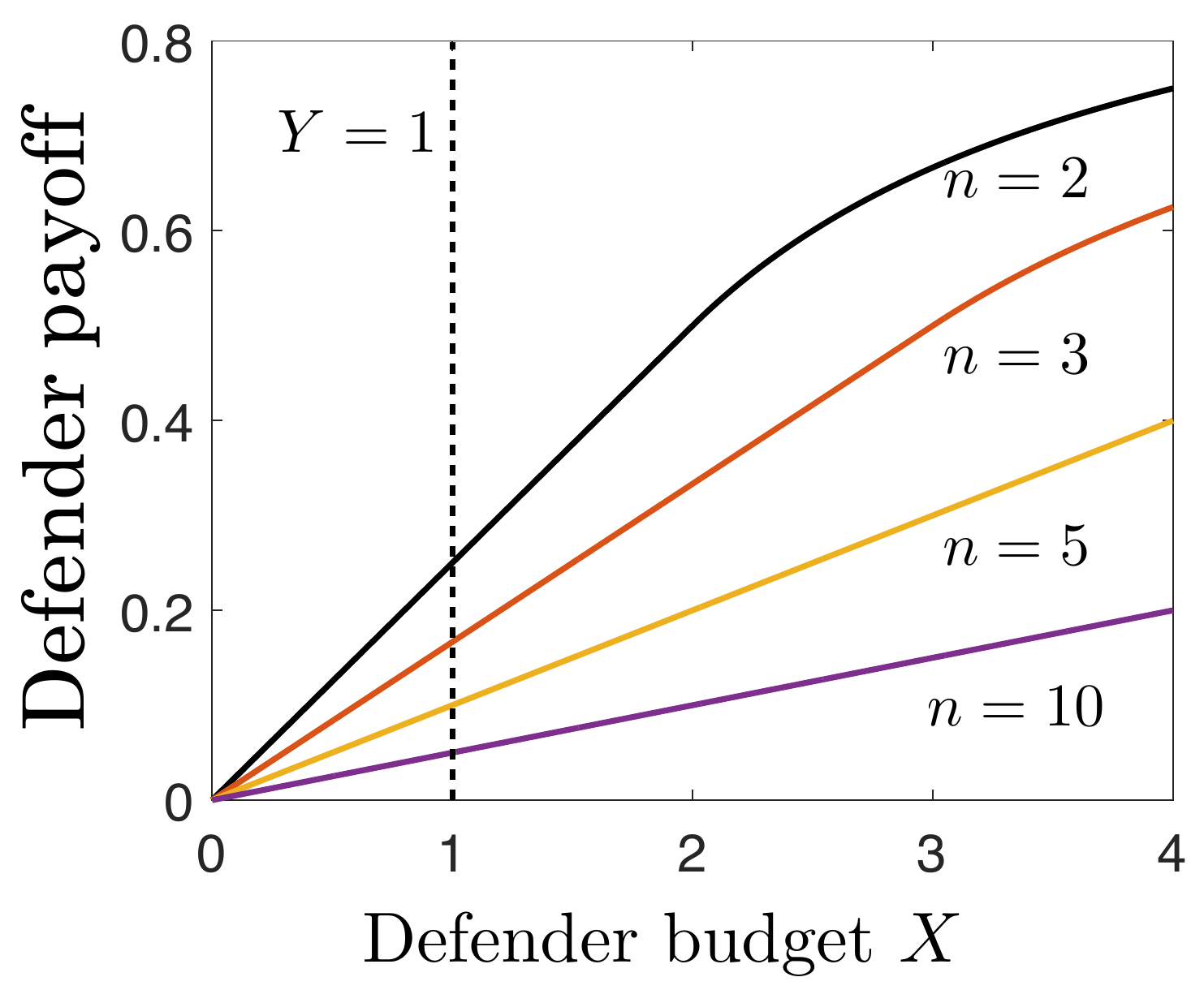}
    \caption{Equilibrium payoff to the defender 
    \eqref{eq:pi_WL} in $\text{WL}(X,Y,\bfs)$ for varying 
    numbers of system vulnerabilities $|\bfs|=2,3,5,10$ with 
    attacker budget fixed at $Y=1$. Maintaining a fixed 
    security probability requires defender resources to scale 
    linearly with $|\bfs|$.}
    \label{fig:WLBS_payoff}
\end{figure}

Figure~\ref{fig:WLBS_payoff} illustrates how vulnerability 
proliferation systematically disadvantages defenders: 
maintaining a fixed security probability requires defender 
resources to scale linearly with the number of contests $|\bfs|$, a challenging 
operational constraint for resource-limited  
operators of large-scale infrastructures. When $X < |\bfs| \cdot Y$, the defender operates 
in a severely resource-constrained regime where breach 
probability exceeds 50\% regardless of strategic choices.

The weakest-link and best-shot objectives of 
Theorem~\ref{thm:WLBS} capture a form of extreme
interdependence in which every battlefield is equally exposed. Recent research has extended this foundation to settings with more complex, localized interdependence structures, 
where system functionality depends on securing specific subsets of battlefields rather than all of them. For a discussion on these threads of research, see  ``\nameref{sidebar-interdependent}''.

\section*{\textbf{Direction 2: From Winner-Take-All to 
Alternate Winning Rules}}

The winning rule considered so far  awards a contest in 
full to whichever player commits more resources. Numerous 
real-world problems cannot be precisely framed using this 
assumption. Valuable contests can be partially split between 
competitors, as in market competition. Outcomes are often 
not determined from a single performance metric such as 
strength, but rather from a combination of factors such as 
quality, reputation, and prior investment. The broader 
contest theory literature has introduced several alternative contest success functions 
(CSFs) $W(x,y): \mathbb{R}^2_{\geq 0} \rightarrow [0,1]$ 
that better capture these nuances, each giving rise to a 
payoff function of the form
\begin{equation}\label{eq:CSF_payoff}
    u_\mcx(\bs{x},\bs{y}) = \sum_{c \in \bfs} v_c \cdot 
    W(x_c, y_c).
\end{equation}
The winner-take-all rule \eqref{eq:blotto_payoff} is 
recovered as a special case. We adopt the convention that $W(x_c, y_c)$ defines the fraction of $v_c$ awarded to player $\mcx$.
In ``\nameref{sidebar-csf}'', we highlight a number of alternative CSFs that have particular relevance to control and engineering applications.

\begin{sidebar}{Contest Success Functions}
\section[Contest Success Functions]{}\label{sidebar-csf}

\setcounter{sequation}{1}
\renewcommand{\thesequation}{S\arabic{sequation}}
\setcounter{stable}{0}
\renewcommand{\thestable}{S\arabic{stable}}
\setcounter{sfigure}{0}
\renewcommand{\thesfigure}{S\arabic{sfigure}}

\sdbarinitial{W}e present a digression to discuss several alternate forms of contest success functions (CSFs) beyond the ``winner-take-all" winning rule that defines the classic Colonel Blotto and General Lotto game, $W(x,y) = \mathds{1}(x_c \geq y_c)$ \eqref{eq:blotto_payoff}. The following examples have been popularized and prominently studied in the contest theory literature \cite{skaperdas_SA,kovenock_2012_SA,vojnovic_SA} .

\subsection{Tullock CSF}

Perhaps the most well-studied is the Tullock CSF. It was initially developed to investigate rent-seeking in economics \cite{Tullock1988}. It is defined as
\begin{sequation}
    W(x,y) = \frac{x}{x+y},
\end{sequation}
which awards player $\mcx$ a fraction of the contest  valuation proportional to its relative resource investment. This fraction is strictly increasing in $x$ with diminishing 
returns, reflecting the intuition that additional investment yields positive but decreasing marginal gains. The Tullock 
CSF is particularly well-suited to model market competition and cybersecurity investment problems where partial rewards and smooth payoff gradients are appropriate 
\cite{Friedman_1958,Osorio_2013,iliaev2023tullock,maljkovic2024blotto,diaz2025strategic}.

\subsection{Lottery CSF}

The lottery CSF is a stochastic 
variant of the Tullock CSF, in which
\begin{sequation}
    W(x,y) = \left\{ \begin{array}{ccl} 1 & & \text{w.p. } 
    \frac{x}{x+y}, \\ 0 & & \text{w.p. } \frac{y}{x+y}. 
    \end{array} \right.
\end{sequation}
Unlike the Tullock CSF, which splits battlefield valuations 
continuously, the lottery CSF preserves the binary winner-take-all outcome structure while introducing probabilistic determination of the winner. This formulation naturally models scenarios where battlefield outcomes are inherently stochastic, such as electronic warfare or anti-jamming, where the success of a defensive countermeasure scales with relative resource investment but remains uncertain in any individual engagement \cite{wu2009optimal,namvar2016jamming,abdelraheem2017cooperative}.

\subsection{Ratio-form CSF}

A systematic study of CSFs was performed by Skaperdas~\cite{skaperdas_SA}, who identified conditions on CSFs to satisfy certain desirable properties.
In particular, CSFs that satisfy a particular set of axioms (which include anonymity, independence of irrelevant alternatives, among other properties), can be written as a \emph{ratio-form CSF},
\begin{sequation}\label{eq:ratio-form}
    W(x,y) = \frac{f(x)}{f(x) + f(y)},
\end{sequation}
where $f: \mbb{R}_+ \rightarrow \mbb{R}_+$ is a continuous and increasing function.
This general form includes many CSFs as special cases, such as the Tullock contest and the logit CSF \cite{dixit1987strategic}.

\subsection{Favoritism}

The favoritism CSF captures 
scenarios where one player holds a structural advantage on certain contests, independent of dynamic resource allocations \cite{vu_SA}. Each contest $c \in \bfs$ is associated with a favoritism parameter $p_c \in \mathbb{R}$ for player $\mcx$, giving rise to the CSF
\begin{sequation}
    W(x_c, y_c; p_c) = \mathds{1}(x_c+p_c\geq y_c).
\end{sequation}
If $p_c > 0$, player $\mcx$ enjoys a home-field advantage on contest $c$; if $p_c < 0$, the advantage belongs to $\mcy$. We note that this is a direct extension to the classic ``winner-take-all" rule, and thus the favoritism CSF has largely appeared in the context of Colonel Blotto and General Lotto games~\cite{washburn2013or,vu_SA,paarporn_SA,magnani2025campaign}. Such asymmetries arise naturally across application 
domains: incumbent political parties hold structural advantages in certain states, established firms maintain loyal customer bases that new entrants must overcome, and 
critical infrastructure operators may have pre-deployed sensing or defensive assets at certain nodes. The favoritism formulation thus bridges static competitive advantage and dynamic resource allocation, enabling analysis of how 
pre-existing positional disparities interact with real-time strategic decisions.

\end{sidebar}

Regardless of which CSF is utilized, the research agenda 
follows the same essential structure from before: characterize equilibrium strategies and payoffs, 
identify how problem parameters shape competitive outcomes, 
and determine what structural properties of optimal policies 
can be extracted to guide practical decision-making. The 
availability of these results across multiple CSF variants 
substantially extends the reach of the framework, enabling 
practitioners to select the formulation whose structural 
assumptions best match the competitive dynamics of their 
application.

\subsection{Strategic pre-allocations via the favoritism CSF}

A particular CSF that is amenable to a deeper analysis is the \emph{favoritism CSF}, defined by 
\begin{equation}\label{eq:fav_CSF}
    % W(x_c, y_c; p_c) = \begin{cases} 1 & \text{if } 
    % x_c + p_c > y_c, \\ \frac{1}{2} & \text{if } 
    % x_c + p_c = y_c, \\ 0 & \text{if } x_c + p_c < y_c
    % \end{cases}
    W(x_c, y_c; p_c) = \mathds{1}(x_c+p_c\geq y_c),
\end{equation}
where $p_c \in \mbb{R}$ denotes the favoritism parameter for player $\mcx$ on contest $c\in\bfs$. Note that $p_c$ can be positive or negative, indicating an inherent advantage or disadvantage on this contest, respectively.

This CSF serves as a natural conceptual bridge towards analyzing scenarios where allocation decisions are made over multiple stages. Specifically, the favoritism parameters $p_c$ need not be exogenously determined. It can 
itself be the outcome of a prior strategic decision, raising 
the question of how a player should optimally invest 
resources to establish competitive positioning before a 
decisive engagement occurs. We address both the static 
equilibrium characterization and this dynamic extension 
in turn.

Favoritism is captured by a vector $\bs{p} \in \mathbb{R}^{|\bfs|}$ 
representing pre-allocated resources across battlefields, 
modifying the payoff to player $\mcx$ as
\begin{equation}\label{eq:fav_payoff}
    u_\mcx(\bs{x},\bs{y}) = \sum_{c \in \bfs} v_c \cdot 
    \mathds{1}\left\{ x_c + p_c \geq y_c \right\},
\end{equation}
with $u_\mcy(\bs{x},\bs{y}) = \phi - 
u_\mcx(\bs{x},\bs{y})$. Here, we recall that $\phi = \sum_c v_c$ denotes the aggregate value of contests. This scenario admits formulations 
as either a Colonel Blotto game  or a General Lotto game \cite{Vu_EC2021}. The 
following result addresses equilibrium characterizations 
for both formulations.

\begin{theorem}[Informal, adapted from \cite{paarporn2024reinforcement} and  \cite{Vu_EC2021}]\label{thm:fav_approx}
    When favoritism 
    is one-sided across all battlefields, that is, $p_c \geq 0$ 
    for all $c \in \bfs$,  the unique equilibrium payoff for the General Lotto formulation can be expressed in closed form. 
    For arbitrary signed $\bs{p}$, 
    there is an algorithm running in time $O(|\bfs|/\delta)$ 
    that produces an $(O(\delta) \cdot \phi)$-equilibrium 
    for the General Lotto formulation, and an 
    $(O(\delta + \gamma) \cdot \phi)$-equilibrium for 
    the Colonel Blotto formulation, where 
    $\gamma = O(1/\sqrt{|\bfs|})$.
\end{theorem}

The equilibrium characterizations above treat $\bs{p}$ as 
a fixed structural feature of the competitive environment. 
In many practical settings, however, a player has the 
opportunity to deliberately engineer this advantage through 
strategic investment prior to the engagement. Consider a 
system operator who, anticipating an adversarial interaction, 
can pre-deploy defensive resources across system components 
to establish favorable competitive positioning before the 
decisive contest begins. Formalizing this intuition leads 
naturally to a two-stage extension in which the favoritism 
vector is itself an endogenous strategic choice.
Indeed, suppose that player $\mcx$ 
has a fixed pre-investment budget $P > 0$ at its disposal.
Such an interaction unfolds as follows.

\vspace{2mm}
\noindent\textbf{Stage 1 (Pre-allocation):} Player $\mcx$ 
selects a pre-allocation vector $\bs{p} \in 
\mathbb{R}^{|\bfs|}_{\geq 0}$ satisfying $\sum_{c \in \bfs} 
p_c = P$. This placement is binding and publicly observed 
by both players.

\vspace{2mm}
\noindent\textbf{Stage 2 (Competition):} Players $\mcx$ and 
$\mcy$ are endowed with resource budgets $X$ and $Y$ 
respectively, and engage in the resulting game with favoritism, 
with final payoffs determined by the equilibrium of this 
stage.
\vspace{2mm}

Player $\mcx$ seeks the pre-allocation $\bs{p}^*$ that 
maximizes its eventual equilibrium payoff  in Stage 2. The 
following theorem characterizes the optimal pre-allocation 
strategy and the resulting payoff.

\begin{theorem}[Adapted from \cite{paarporn2024reinforcement}]
\label{thm:GLP} 
    Consider the General Lotto formulation of the two-stage extension. The optimal pre-allocation vector for player $\mcx$ is 
    \be
        p_c^* = v_c \cdot P.
    \ee
    The resulting payoff 
    to player $\mcx$ in Stage 2 is determined according to three cases.
    \begin{enumerate}
        \item If $Y \leq P$, or $Y > P$ and $X \geq 
        \frac{2(Y-P)^2}{P+2(Y-P)}$, then the payoff is
        \begin{equation*}
            \phi\cdot\left( 1 - \frac{Y}{2X}
            \left(\frac{X + \sqrt{X(X+2P)}}{P + X + 
            \sqrt{X(X+2P)}} \right)^2 \right).
        \end{equation*}
        \item If $Y > P$ and $0 < X < 
        \frac{2(Y-P)^2}{P+2(Y-P)}$, then 
        the payoff is $\phi\cdot\frac{X}{2(Y-P)}$.
        \item If $X = 0$, then the payoff is $ 
        \phi\cdot(1 - \min\{\frac{Y}{P},1 \})$.
    \end{enumerate}
    From the constant-sum structure, player $\mcy$'s resulting payoff is the payoff for player $\mcx$ subtracted from $\phi$.
\end{theorem}

\begin{figure}
    \centering
    \includegraphics[width=1.0\linewidth]{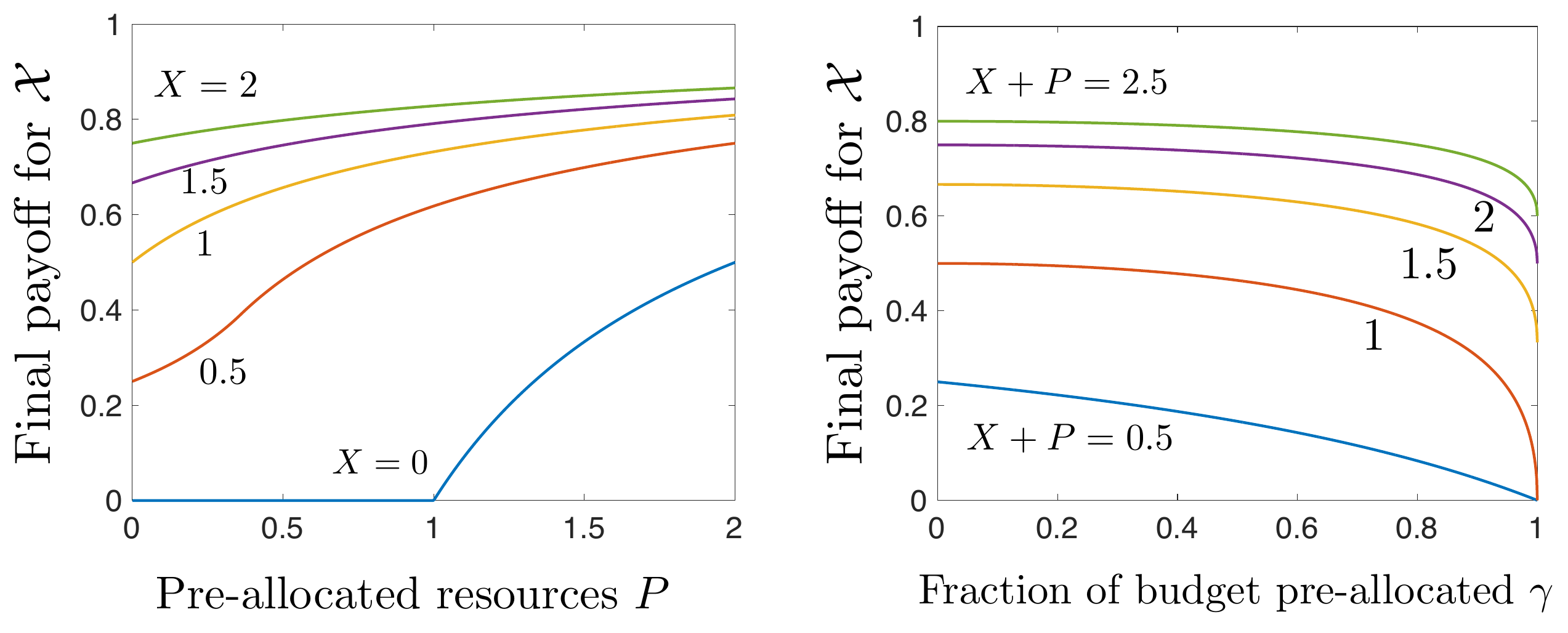}
    \caption{(Left) This plot shows the final payoff to player $\mcx$ in Stage 2 of the two-stage extension under the optimal pre-allocation of $P$ resources, identified in Theorem~\ref{thm:GLP}. In this plot, the budget for player $\mcy$ is fixed at $Y=1$. 
    (Right) This plot shows the final payoff to player $\mcx$ when it has a fixed \emph{total budget} (pre-allocated and regular resources), but the composition of this endowment varies from having no pre-allocated resources ($\gamma = 0$), to having only pre-allocated resources ($\gamma=1$).
    When only pre-deployed resources are available, a sufficient amount is needed to obtain a positive final payoff. Overall, the payoff is strictly decreasing as the composition of budget contains more pre-allocated resources. This is because the allocation of pre-deployed resources is public by the time of the Stage 2 competition, indicating that, per unit of budget, the pre-allocated resources ($P$) are less effective than regular resources ($X$). }\label{fig:favoritism_plots}
\end{figure}

A few remarks are in order. First, the optimal pre-allocation 
distributes $P$ resources proportionally to the battlefield 
valuations, and as $P \to 0$ the resulting payoffs reduce 
continuously to those of the classic General Lotto game 
\eqref{eq:lotto_equil}. Second, the closed-form expressions 
above constitute precise analytical characterizations, made 
possible by restricting to one-sided favoritism across all 
battlefields. When $\bs{p}$ has arbitrary signed values, exact 
equilibria remain elusive and the computational approaches 
of Theorem~\ref{thm:fav_approx} represent the sharpest 
available characterizations. Third, the payoffs exhibit a 
nontrivial dependence on all three budget parameters $P$, 
$X$, and $Y$, the structure of which is illustrated in 
Figure~\ref{fig:favoritism_plots} for a range of parameter configurations.

%\jason{I think we need to really assess what formalisms for game notation are needed.  I know we need GL(...) but not sure about all the variants.}

\section*{\textbf{Direction 3: Multi-agent Competitive 
Environments}}

The models examined thus far share a common structural 
assumption: two players competing directly against one 
another. While this one-versus-one framework captures the 
essential tension of adversarial resource allocation, it 
omits a class of strategic considerations that arise 
naturally in practice. 
Industrial 
control operators, autonomous vehicle fleets, and multi-robot 
systems all face scenarios where multiple decision-makers 
simultaneously contend with a shared adversary. 
% Real-world competitive environments 
% rarely reduce to isolated bilateral contests: industrial 
% control operators, autonomous vehicle fleets, and multi-robot 
% systems all face scenarios where multiple decision-makers 
% simultaneously contend with a shared adversary. 
In such 
settings, the strategic question expands well beyond how to 
allocate resources across contests. It encompasses many other strategic mechanisms:  whether to form alliances, to reveal information to certain competitors, or to transfer competitive assets to other players.
% whether to form alliances, how much information to reveal, 
% and whether publicly conceding competitive assets can 
% paradoxically improve outcomes. 
These richer mechanisms, 
which have no analog in the two-player setting, motivate extensions to multi-agent 
competitive environments.

\subsection{The Coalitional Blotto Game}

\begin{figure}
    \centering
    \includegraphics[width=1.0\linewidth]{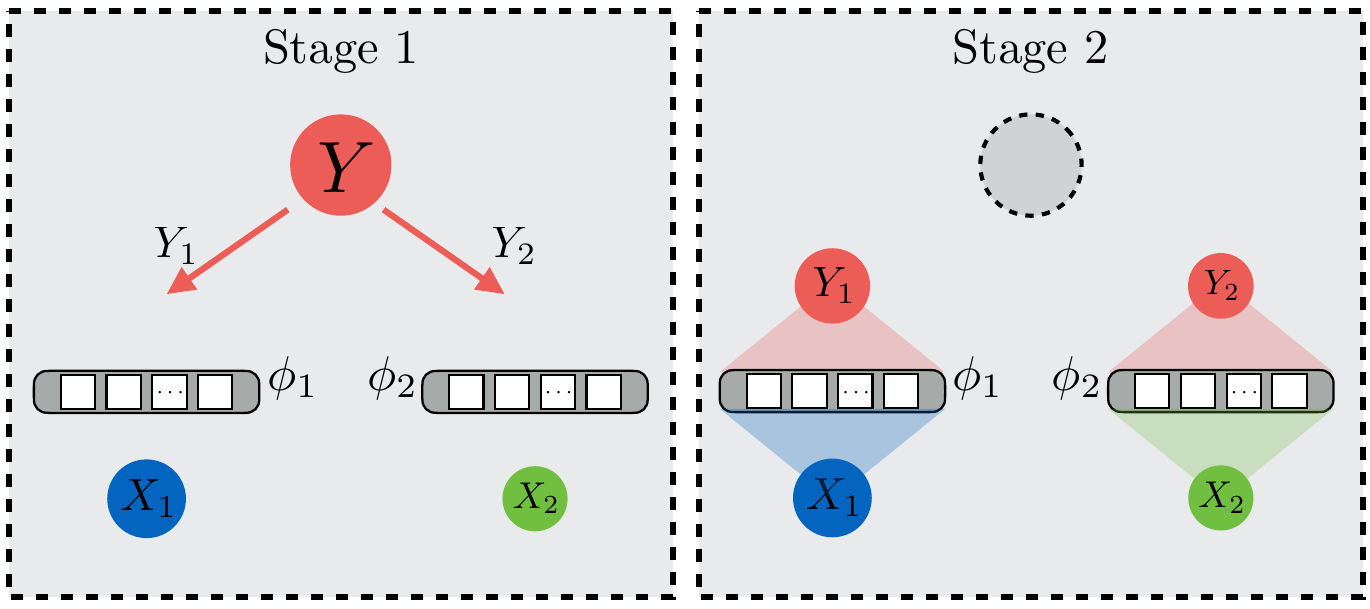}
    \caption{Illustration of the Coalitional Blotto game. Players $\mcx_1$ 
    and $\mcx_2$ compete against a common opponent $\mcy$ on 
    separate and disjoint sets of contests, using resource 
    budgets $X_1$, $X_2$, and $Y$ respectively. The contest 
    between $\mcx_i$ and $\mcy$ has total value $\phi_i$. The 
    opponent divides its budget between the two battlefield 
    sets in Stage 1, after which two independent General Lotto 
    games are played in Stage 2.}
    \label{fig:GL3}
\end{figure}

The Coalitional Blotto game, first proposed in \cite{Kovenock_2012}, provides a canvas for 
studying these richer strategic interactions. Here, two 
players $\mcx_1$ and $\mcx_2$ simultaneously compete against 
a common opponent $\mcy$, with resource budgets $X_1$, $X_2$, 
and $Y$, respectively. Player $\mcx_1$ uses its resources 
to compete with $\mcy$ over a set of contests $\bfs_1$, 
and player $\mcx_2$ uses its resources to compete with $\mcy$ 
over a disjoint set of contests $\bfs_2$. The aggregate value 
of contests in $\bfs_i$ is denoted $\phi_i > 0$, 
$i = 1, 2$. Figure~\ref{fig:GL3} provides an illustration 
of this scenario. The central strategic decision for $\mcy$ 
is how to divide its budget across the two separate and disjoint
contest sets. For each, a subordinate agent $\mcy_i$ 
($i=1,2$) is endowed with resources $Y_i$ and acts on behalf 
of $\mcy$, engaging player $\mcx_i$ in a General Lotto game 
$\text{GL}(X_i,Y_i,\phi_i)$. The Coalitional Blotto game unfolds in two 
stages.

\vspace{2mm}

\begin{framed}

\noindent\textbf{Stage 1:} Opponent $\mcy$ selects a 
division $(Y_1,Y_2)$ satisfying $Y_1 + Y_2 = Y$. Subordinate 
agents $\mcy_1$ and $\mcy_2$ are then endowed with resources 
$Y_1$ and $Y_2$, respectively. The division is publicly observable to all agents.

\vspace{2mm}
\noindent\textbf{Stage 2:} Two independent General Lotto 
games are simultaneously played: $\text{GL}(X_1,Y_1,\phi_1)$ 
between $\mcx_1$ and $\mcy_1$, and $\text{GL}(X_2,Y_2,
\phi_2)$ between $\mcx_2$ and $\mcy_2$.
\end{framed}

\vspace{2mm}

In Stage 2, we assume that players employ equilibrium strategies in their respective General Lotto games. From Theorem~\ref{thm:GL}, the subsequent payoff that agent 
$\mcx_i$ obtains in Stage 2 is
\begin{equation}\label{eq:GL3_pi_i}
    \pi_i(X_i,Y_i) := \phi_i \cdot L(X_i,Y_i),
\end{equation}
and the payoff to $\mcy$ is the sum of its subordinates' 
payoffs:
\begin{equation}
    \pi_{\mcy}(Y_1,Y_2;\Gamma) := (\phi_1 - \pi_1(X_1,Y_1)) 
    + (\phi_2 - \pi_2(X_2,Y_2)),
\end{equation}
where we refer to $\Gamma := (X_1,X_2,\phi_1,\phi_2)$ as the \emph{game instance}, or the tuple specifying all of the game's parameters.  

The opponent $\mcy$ seeks to maximize its total payoff across the two engagements, and consequently must find an optimal division,
\begin{equation}\label{eq:opt_division}
    (Y_1^*(\Gamma),Y_2^*(\Gamma)) \in \arg\max_{(Y_1,Y_2)} 
    \pi_{\mcy}(Y_1,Y_2;\Gamma),
\end{equation}
This optimization problem was solved analytically in 
\cite{Kovenock_2012}. Under the opponent's optimal division, the final payoffs are expressed as
\begin{equation}
    \pi_i^*(\Gamma) := \pi_i(X_i,Y_i^*(\Gamma)), \quad 
    i=1,2.
\end{equation}
With this foundation in place, we now turn to the central 
question: what strategic mechanisms can $\mcx_1$ and $\mcx_2$ 
exploit to improve their individual performances against the 
common opponent?

\subsection{Alliance formation through budget transfers}

A strategic mechanism we examine is a budgetary 
transfer between $\mcx_1$ and $\mcx_2$. Before 
Stage 1 unfolds, suppose $\mcx_1$ and $\mcx_2$ 
agree on a transfer $t \in [-X_2, X_1]$ of resources between 
them, where $t > 0$ represents a transfer from $\mcx_1$ to 
$\mcx_2$ and $t < 0$ represents the reverse. It is important 
to emphasize that each player is solely concerned with the 
outcomes on their own set of contests: $\mcx_i$ cares 
only about its payoff $\pi_i^*$ and is indifferent to 
outcomes on $\bfs_{-i}$. A transfer therefore constitutes 
a meaningful sacrifice: a player voluntarily weakens its own 
competitive position in exchange for the possibility of a 
more favorable response from $\mcy$. This motivates 
appending a Stage 0 that occurs prior to Stage 1:

\smallskip

\begin{framed}

\noindent\textbf{Stage 0:} Players $\mcx_1$ and $\mcx_2$ 
jointly select a transfer $t \in [-X_2,X_1]$. Their 
resulting resource budgets become $X_1 - t$ and $X_2 + t$, 
respectively. The transfer is publicly observed by $\mcy$ 
prior to its division decision in Stage 1.
\end{framed}

\smallskip

\begin{figure}
    \centering
    \includegraphics[width=1.0\linewidth]{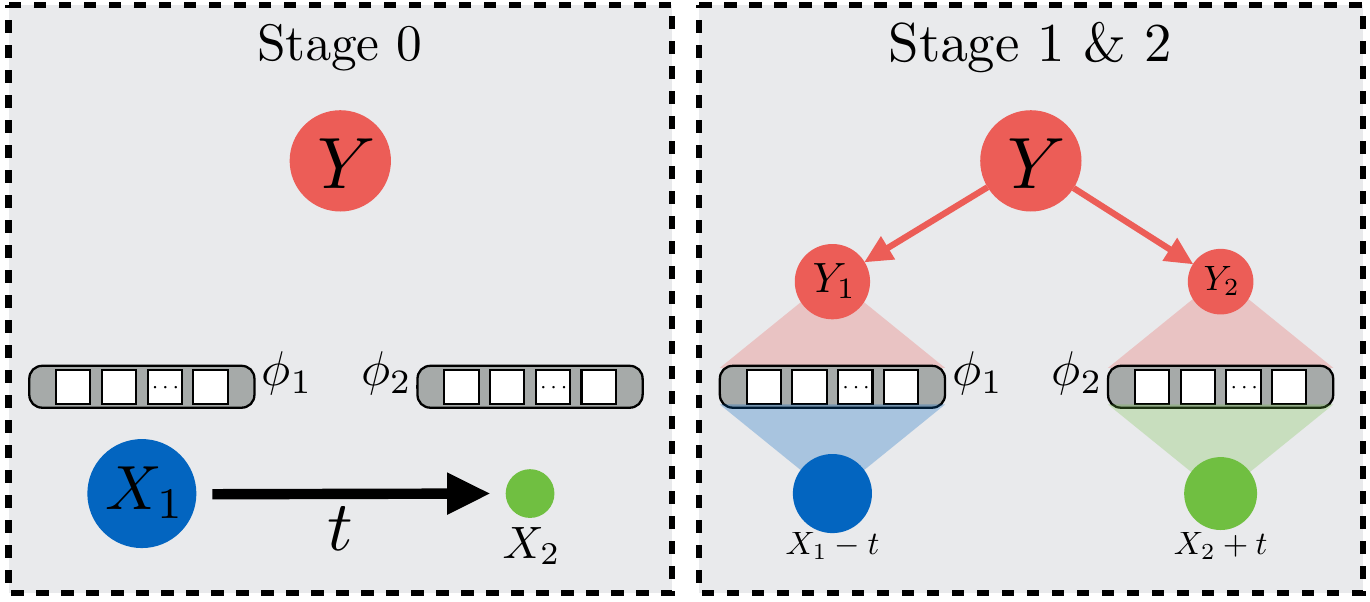}
    \caption{The coalitional Blotto game with budget 
    transfers. In Stage 0, player $\mcx_1$ transfers an 
    amount $t$ from its budget to player $\mcx_2$. The 
    transfer is public knowledge. Then, the three-player 
    Blotto game unfolds in Stages 1 and 2.}
    \label{fig:GL3-transfers}
\end{figure}

Figure~\ref{fig:GL3-transfers} illustrates the full sequence of this interaction. The 
natural question is whether there exist transfers that 
strictly improve both players' payoffs simultaneously. That is, 
whether there exists a $t$ such that
\begin{equation}
    \pi_i^*(\Gamma_t) > \pi_i^*(\Gamma), \quad i = 1,2,
\end{equation}
where $\Gamma_t := (X_1-t, X_2+t, \phi_1, \phi_2)$. When 
such a condition holds, we say that $\mcx_1$ and $\mcx_2$ 
are able to form a mutually beneficial alliance.

\begin{theorem}[Informal, adapted from \cite{Kovenock_2012}]\label{thm:coalition}
There exists a positive measure set of game instances in 
which a mutually beneficial alliance between $\mcx_1$ and 
$\mcx_2$ can be formed through a budgetary transfer, 
strictly improving both players' final Stage 2 payoffs 
simultaneously.
\end{theorem}

The implications of this result are striking. A player voluntarily reduces its own budget through a transfer, yet its equilibrium payoff strictly improves. The mechanism is indirect: the transfer shifts $\mcy$'s attention toward the receiving player's contests, reducing the competitive pressure on the transferring player's own contest set. The net effect, reduced adversarial pressure combined with 
the strategic reallocation of resources, yields a strict improvement for both players simultaneously. This fundamentally challenges the intuition that competitive strength is monotone in resources, and illustrates how 
strategic interdependence can create opportunities that are invisible from a purely optimization-based perspective.

Follow-up research has examined alternative strategic mechanisms that, unlike alliance formation, does not require a mutual agreement between parties to enact~\cite{shah2024aamas,chandanArt2025}.
The mechanisms studied are called \emph{concessions}, in which a player unilaterally and voluntarily reduces its own competitive assets. A summary of these formulations and the research findings are detailed in ``\nameref{sidebar-concessions}''.

\begin{sidebar}{The Art of Concessions}
\section[The Art of Concessions]{}\label{sidebar-concessions}

\setcounter{sequation}{5}
\renewcommand{\thesequation}{S\arabic{sequation}}
\setcounter{stable}{0}
\renewcommand{\thestable}{S\arabic{stable}}
\setcounter{sfigure}{3}
\renewcommand{\thesfigure}{S\arabic{sfigure}}

\sdbarinitial{T}he alliance mechanism represented with budget transfers requires a mutual agreement between $\mcx_1$ 
and $\mcx_2$. In many practical settings, however, the coordination channels necessary to negotiate and execute such agreements may not be available due to physical, institutional, or strategic constraints. This raises a natural question: are there \emph{unilateral} mechanisms for which a single player can take independently, without the consent of any other party, that can improve its own competitive position? Still within the context of the Coalitional Blotto game, we examine two such unilateral mechanisms based on \emph{concessions}, in which a player deliberately weakens itself.

\subsection{Budget concessions}

Consider a player that voluntarily removes a 
portion of its own resources from competition entirely, publicly reducing its effective budget. Specifically, the Stage 0 that takes place before the Coalitional Blotto game unfolds is defined as follows.

\vspace{1mm}

\noindent\textbf{Stage 0 (Budget Concession):} Player $\mcx_1$ selects an amount $x_1 \in [0, X_1]$ to remove from its budget. Its effective budget becomes $X_1 - x_1$. The action is binding and publicly observed by all other players.

\vspace{1mm}

\sdbarfig{\includegraphics[width=18.0pc]{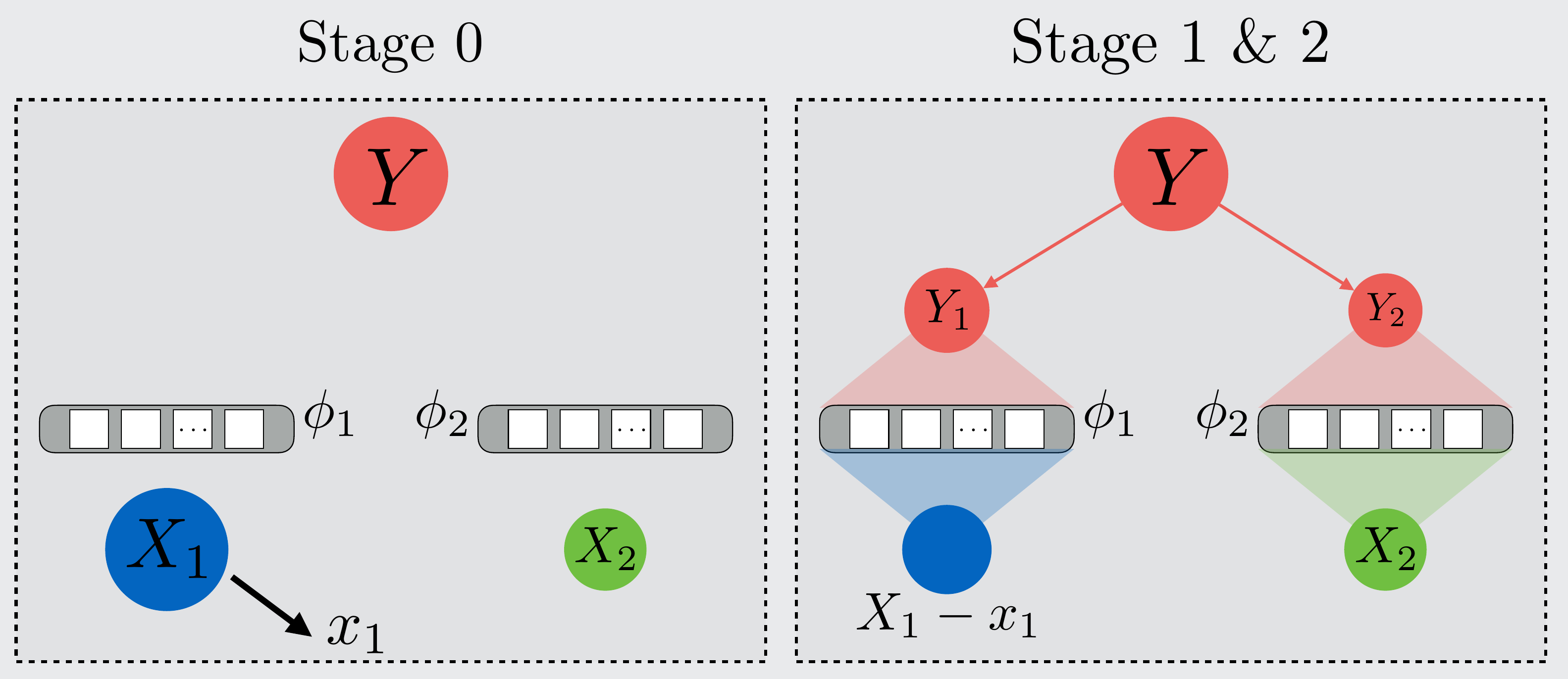}}{Budget concessions. Player $\mcx_1$ removes some amount of its original budget in Stage 0 before engaging in the coalitional Blotto game in Stages 1 and 2.\label{sfig-budgetconcession}}

Figure~\ref{sfig-budgetconcession} illustrates the full sequence of this interaction. A budget concession $x_1 \in (0, X_1]$ is beneficial if 
$\pi_1^*(\Gamma(x_1)) > \pi_1^*(\Gamma)$, where 
$\Gamma(x_1) := (X_1 - x_1, X_2, \phi_1, \phi_2)$ denotes modified parameters of the subsequent Coalitional Blotto game. The 
following result establishes that no such concession exists.

\begin{theorem}[Informal, adapted from \cite{chandan_2025_SA}]\label{thm:budget_concession}
There is no game instance for which a budget concession is 
beneficial to the conceding player.
\end{theorem}

This result confirms a natural intuition: unilaterally discarding resources cannot improve one's competitive position.
Interestingly, this conclusion does not 
extend to all forms of concession.

\subsection{Value concessions}

In a \emph{value concession}, a player instead surrenders a portion of its contested contest value directly to the opponent. Specifically, the Stage 0 is amended as follows.

\vspace{1mm}

\noindent\textbf{Stage 0 (Value Concession):} Player $\mcx_1$ selects an amount $v_1 \in [0, \phi_1]$ of its contest value to surrender to $\mcy$. The opponent $\mcy$ is immediately awarded this value, and the contest between $\mcx_1$ and $\mcy$ is now valued at $\phi_1 - v_1$. The action is binding and publicly observed by all other players.

\vspace{1mm}

\sdbarfig{\includegraphics[width=18.0pc]{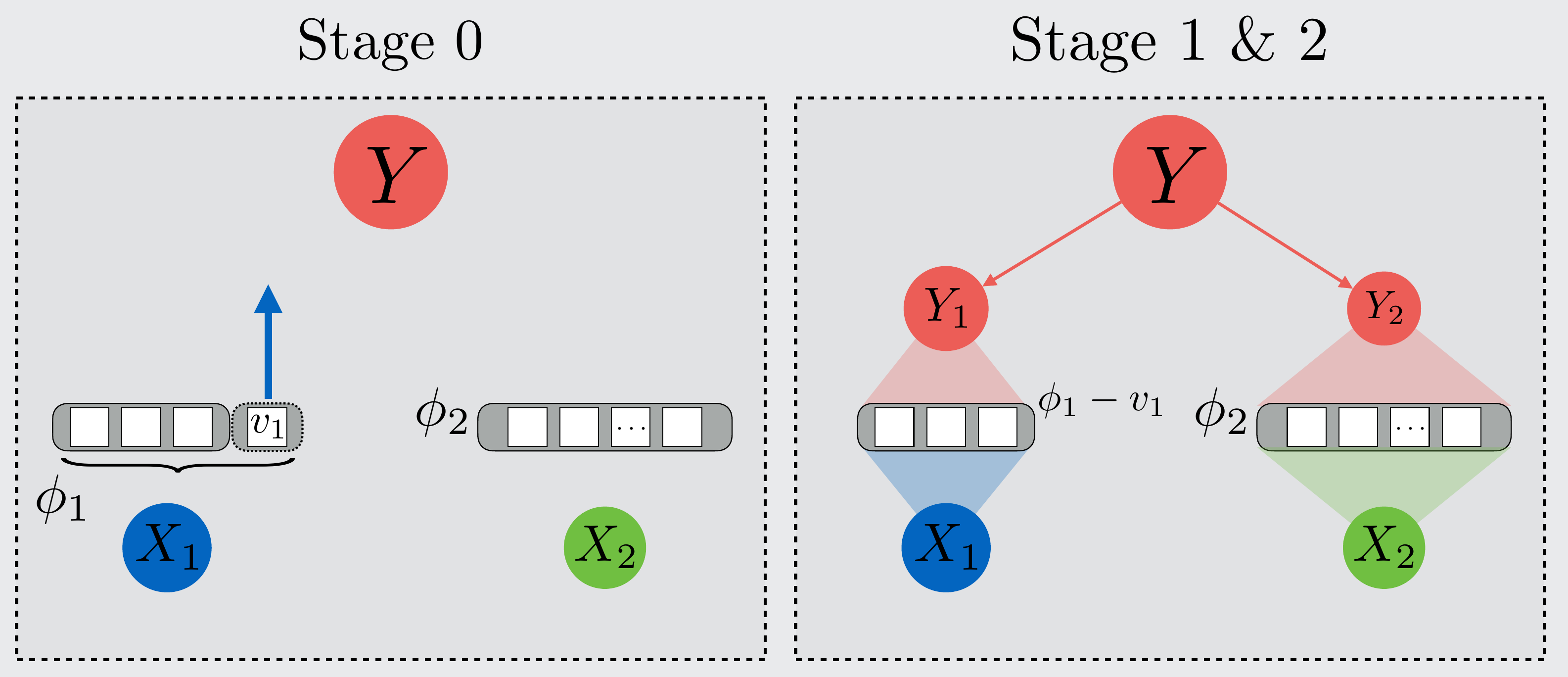}}{Value concessions. Player $\mcx_1$ surrenders a portion $v_1 \in [0,\phi_1]$ of its battlefield value to $\mcy$ in Stage 0, before engaging in the Coalitional Blotto game in Stages 1 and 2.\label{sfig-valueconcession}}

Figure~\ref{sfig-valueconcession} illustrates the full sequence of this interaction. A value concession $v_1 \in (0, \phi_1]$ is beneficial to agent $\mcx_1$ if $\pi_1^*(\Gamma(v_1)) > \pi_1^*(\Gamma)$, where 
$\Gamma(v_1) := (X_1, X_2, \phi_1-v_1, \phi_2)$. In contrast to budget concessions, the following result establishes that beneficial value concessions can indeed exist.

\begin{theorem}[Informal, adapted from \cite{chandan_2025_SA}]\label{thm:value_concession}
There exists a positive measure set of game instances for which a value concession is beneficial to the conceding player.
\end{theorem}

The contrast with the budget concession result is sharp. Surrendering resources unconditionally never helps, yet surrendering contest valuations to the common adversary can. The mechanism mirrors that of the budgetary transfer: by reducing the attractiveness of its own battlefields, $\mcx_1$ induces $\mcy$ to redirect a greater share of its budget toward $\mcx_2$, relieving competitive pressure on $\mcx_1$'s remaining contest. The reduction in contest value is therefore not purely a loss; it is a strategic signal that reshapes the opponent's allocation decision in a favorable direction.

\end{sidebar}

Together, these results illustrate that multi-agent competitive 
environments give rise to a qualitatively richer class of 
strategic considerations than those present in two-player competitive settings. Beyond the allocation of resources across contests, players must reason about coalition formation, 
unilateral concessions, and the strategic signals embedded in their own actions. This complexity has motivated a growing body of research examining additional mechanisms through which players can influence competitive outcomes, including the strategic value of revealing intentions prior to 
competition~\cite{Chandan_2020,paarporn2024strategically}, the transfer of contest responsibilities between allied 
players \cite{shah2024aamas}, and the role of informational asymmetries in shaping coalition incentives \cite{Gupta_2014b}. Taken together, these directions reflect a broader recognition that in adversarial multi-agent 
settings, competitive advantage is not determined by resource levels alone; it is shaped equally by the strategic architecture of the engagement itself.

\section{Concluding Remarks}

This article has presented Colonel Blotto games as a 
unifying framework for competitive resource allocation 
problems that are directly relevant to the controls 
community. The classic formulation captures the essential 
tension of adversarial allocation in its simplest form: two 
budget-constrained players simultaneously distributing 
resources across multiple contested objectives, with 
equilibrium strategies providing worst-case performance 
guarantees that parallel the robustness objectives familiar 
from $\mathcal{H}_\infty$ control. The General Lotto 
relaxation makes this framework analytically tractable, 
admitting closed-form equilibrium characterizations for 
arbitrary parameter configurations and serving as the 
foundation for the three directions surveyed here.

Each of those directions addresses a structural limitation 
of the classic formulation that is particularly consequential 
for engineered systems. Interdependent contest objectives 
capture the all-or-nothing vulnerabilities of networked 
infrastructure, where a single undefended node can 
compromise an entire system. Alternate winning rules, and 
the favoritism formulation in particular, extend the 
framework to settings with partial rewards, stochastic 
outcomes, and pre-existing competitive asymmetries that 
model realistic operational environments. Multi-agent 
competitive environments reveal that in settings with 
multiple decision-makers contending with a shared adversary, 
competitive advantage is shaped not only by resource levels 
but by the strategic architecture of the engagement itself: 
alliances, concessions, and the deliberate revelation of 
intentions can all shift equilibrium outcomes in ways that 
are invisible from a purely optimization-based perspective.

Despite the substantial progress surveyed here, many 
directions remain open. Analytical equilibrium 
characterizations for the most general Colonel Blotto 
formulations, combining asymmetric budgets with 
heterogeneous and arbitrary battlefield valuations, remain 
elusive. Extensions to settings with dynamic and sequential 
decision-making, incomplete information about the opponent's 
resources or objectives, and more than two competing parties 
are active research frontiers. For the controls community 
in particular, the integration of Colonel Blotto frameworks 
with learning-based and data-driven approaches represents 
a promising avenue, enabling adaptive allocation strategies 
in environments where the adversary's behavior is unknown 
or time-varying. The breadth of open problems, combined 
with the framework's demonstrated applicability across 
cybersecurity, network defense, and multi-agent systems, 
suggests that Colonel Blotto games will remain a productive 
and growing area of research at the intersection of game 
theory and control systems for years to come.

\section{ACKNOWLEDGMENT}

This work is partially supported by AFOSR grants \#FA9550-25-1-0245 and \#FA9550-21-1-0203, NASA grant \#103215, and NSF grant \#ECCS-2346791.

\section{Author Information}

\begin{IEEEbiography}{{K}eith Paarporn}{\,}(kpaarpor@uccs.edu) received the B.S. degree from the University of Maryland, College Park in 2013, the M.S. in Electrical and Computer Engineering from the Georgia Institute of Technology in 2016, and a Ph.D. in Electrical and Computer Engineering from the Georgia Institute of Technology in 2018. From 2018 to 2022, he was a postdoctoral scholar in the Electrical and Computer Engineering Department at the University of California, Santa Barbara. He is currently an Assistant Professor in the Department of Computer Science at the University of Colorado, Colorado Springs. He is the recipient of an Engineering Research Initiation (ERI) Award from the NSF in 2024, and an NSF CAREER Award in 2026. His research interests include game theory, control theory, and their applications to multi-agent systems and security. He is a Member of IEEE. 
\end{IEEEbiography}

\begin{IEEEbiography}{Jason R. Marden}{\,}(jrmarden@ece.ucsb.edu) received the B.S. in Mechanical Engineering in 2001 from UCLA, and a Ph.D. in Mechanical Engineering in 2007, also from UCLA. After graduating from UCLA, he served as a junior fellow in the Social and Information Sciences Laboratory at the California Institute of Technology until 2010 when he joined the University of Colorado. In 2015, Jason joined the Department of Electrical and Computer Engineering at UCSB.  Jason is a recipient of the NSF Career Award (2014), the ONR Young Investigator Award (2015), the AFOSR Young Investigator Award (2012), the American Automatic Control Council Donald P. Eckman Award (2012), the SIAG/CST Best SICON Paper Prize (2015), and was named an IEEE Fellow (2023). Jason’s research interests focus on game theoretic methods for the control of distributed multiagent systems.
\end{IEEEbiography}

% % % % % % % % % % % % % % % % % % % %

\bibliographystyle{IEEEtran}
\bibliography{sources}

\endarticle

\end{document}